\documentclass[%
 reprint,
showkeys,
 amsmath,amssymb,
 aps,
pre
]{revtex4-2}

\usepackage{graphicx}
\usepackage{dcolumn}
\usepackage{bm}

\usepackage{nicefrac}
\usepackage{makecell} 		
\usepackage{rotating}					
\newcommand{\0}{\text{0}}
\newcommand{\1}{\text{1}}
\newcommand{\2}{\text{2}}
\newcommand{\3}{\text{3}}
\newcommand{\4}{\text{4}}
\newcommand{\6}{\text{6}}
\newcommand{\comma}{\text{,}}

\usepackage{accents}
\newcommand*{\dt}[1]{%
  \accentset{\mbox{\large\bfseries .}}{#1}}

\begin{document}


\title{A Rheological Analogue for Brownian Motion with Hydrodynamic Memory}

\author{Nicos Makris}
 \email{nmakris@smu.edu}
 \altaffiliation[Also at ]{Office of Theoretical and Applied \\Mechanics, Academy of Athens, 10679, Greece}
\affiliation{%
 Dept. of Civil and Environmental Engineering, Southern Methodist University, Dallas, Texas, 75276
}%


\date{\today}

\begin{abstract}
When the density of the fluid surrounding suspended Brownian particles is appreciable, in addition to the forces appearing in the traditional Ornstein and Uhlenbeck theory of Brownian motion, additional forces emerge as the displaced fluid in the vicinity of the randomly moving Brownian particle acts back on the particle giving rise to long-range force correlations which manifest as a ``long-time tail'' in the decay of the velocity autocorrelation function known as hydrodynamic memory. In this paper, after recognizing that for Brownian particles immersed in a Newtonian, viscous fluid, the hydrodynamic memory term in the generalized Langevin equation is essentially the $\nicefrac{\text{1}}{\text{2}}$ fractional derivative of the velocity of the Brownian particle, we present a rheological analogue for Brownian motion with hydrodynamic memory which consists of a linear dashpot of a fractional Scott-Blair element and an inerter. The synthesis of the proposed mechanical network that is suggested from the structure of the generalized Langevin equation simplifies appreciably the calculations of the mean-square displacement and its time-derivatives which can also be expressed in terms of the two-parameter Mittag--Leffler function.
\end{abstract}

\maketitle


\section{Introduction}

Following \citeauthor{Einstein1905}'s \cite{Einstein1905} `long-term' theory of Brownian motion, Langevin \cite{Langevin1908} presented a force--balance dynamic approach where the inertia of a Brownian microsphere with mass $m$ and radius $R$ balances the friction force that develops as the particle moves inside the surrounding fluid, together with the random excitation forces $f_R(t)$, that originate from the collisions of the fluid molecules on the Brownian microsphere
\begin{equation}\label{eq:Eq01}
m\frac{\mathrm{d}v(t)}{\mathrm{d}t}=-\zeta v(t)+f_R(t)
\end{equation}
In Eq. \eqref{eq:Eq01}, $v(t)=\frac{\mathrm{d}r(t)}{\mathrm{d}t}$ is the particle velocity and $\zeta v(t)$ is a viscous drag force proportional to the velocity of the Brownian particle. For a steady motion in a memoryless, viscous fluid with viscosity $\eta$, the drag coefficient is given by Stokes' law $\zeta=\6\pi R\eta$ \citep{LandauLifshitz1959} which has been traditionally accepted when long-range correlations are not present \citep{Attard2012, KalmykovCoffey2017}. According to the Ornstein and Uhlenbeck theory of Brownian motion \citep{UhlenbeckOrnstein1930,WangUhlenbeck1945}, the random excitation $f_R(t)$ has a zero average value over time $\left\langle f_R(t) \right\rangle=$ 0; while for the memoryless, Newtonian viscous fluid that only dissipates energy, the correlation function contracts to a Dirac delta function \citep{Lighthill1958}
\begin{equation}\label{eq:Eq02}
\left\langle f_R(t_\1) f_R(t_\2) \right\rangle=A \delta(t_\1-t_\2)
\end{equation}
with $A$ being a constant that expresses the strength of the random forces. Equation \eqref{eq:Eq02} indicates that successive collisions of the fluid molecules on the Brownian particle are uncorrelated.

Given the random nature of the excitation force $f_R(t)$, the Langevin Eq. \eqref{eq:Eq01} can be integrated in terms of ensemble averages in association with Eq. \eqref{eq:Eq02} and the velocity autocorrelation function of Brownian particles immersed in a memoryless viscous, Newtonian fluid with viscosity $\eta$ is \citep{UhlenbeckOrnstein1930, WangUhlenbeck1945}
\begin{align}\label{eq:Eq03}
\left\langle v(\0) v(t) \right\rangle &=\left\langle v(\xi) v(\xi+t) \right\rangle \\ \nonumber
& = \lim_{T \rightarrow \infty} \frac{\1}{T} \int_\0^T v(\xi) v(\xi+t) \, \mathrm{d}\xi = \frac{N K_B T}{m} e^{-\nicefrac{t}{\tau}}
\end{align}
where $N \in \left\lbrace \1 \comma \2 \comma \3 \right\rbrace$ is the number of spatial dimensions, $K_B$ is Boltzmann's constant, $T$ is the equilibrium temperature and $\tau=\frac{m}{\6\pi R \eta}$ is the dissipation time-scale of the perpetual fluctuation--dissipation process.

When the density of the fluid surrounding the Brownian particle is appreciable, there is a convoluted interplay between the unsteady motion of the particle and the motion of the displaced fluid. Accordingly, in addition to the forces appearing in Eq. \eqref{eq:Eq01}, the Brownian particle also displaces the fluid in its immediate vicinity and in return the fluid acts back on the particle and gives rise to long-range correlations which are different than the delta-correlations of Eq. \eqref{eq:Eq02}. The interaction of the Brownian particle with the displaced fluid can be described by the addition of inertia and memory terms to the basic Langevin Eq. \eqref{eq:Eq01} \citep{LandauLifshitz1959, ZwanzigBixon1970, Widom1971, Hinch1975, PaulPusey1981, ClercxSchram1992}.
\begin{align}\label{eq:Eq04}
m\frac{\mathrm{d}v(t)}{\mathrm{d}t} & =-\frac{\2}{\3}\pi R^\3 \rho_f \frac{\mathrm{d}v(t)}{\mathrm{d}t} - \6\pi R \eta v(t) \\ \nonumber 
& - \6 R^\2 \sqrt{\pi \rho_f \eta} \int_{\0^-}^t \frac{\nicefrac{\mathrm{d}v(\xi)}{\mathrm{d}\xi}}{(t-\xi)^{\nicefrac{\1}{\2}}}\, \mathrm{d}\xi + f_R(t)
\end{align}
The quantity $\frac{\2}{\3}\pi R^\3 \rho_f$ in the additional inertia term of Eq. \eqref{eq:Eq04} is half the mass of the displaced fluid $\frac{m_f}{\2}$ $\bigg( m_f=\frac{\4}{\3}\pi R^\3 \rho_f$ with $\rho_f=$ density of the surrounding fluid$\bigg)$; whereas, the following term $\6\pi R \eta v(t)$ is the ordinary Stokes' friction force, also present in Eq. \eqref{eq:Eq01}. The convolution integral term in Eq. \eqref{eq:Eq04} is a hydrodynamic memory force that emerges from the reaction of the displaced fluid on the moving Brownian particle \citep{LandauLifshitz1959}.

By recalling that the Riemann--Liouville fractional integral of a continuous function $f(t)$ is defined as \citep{OldhamSpanier1974, SamkoKilbasMarichev1974, MillerRoss1993, Podlubny1998}
\begin{equation}\label{eq:Eq05}
I^q\left[ f(t) \right] = \frac{\1}{\Gamma(q)}\int_{\0^-}^t \frac{f(\xi)}{(t-\xi)^{\1-q}} \, \mathrm{d}\xi \comma \quad q\in \mathbb{R}^+
\end{equation}
the hydrodynamic memory integral in Eq. \eqref{eq:Eq04} is essentially the fractional integral of order $q=$ {\large $\nicefrac{\1}{\2}$} of the acceleration history $\frac{\mathrm{d}v(t)}{\mathrm{d}t}$ of the Brownian particle. Accordingly, by virtue of Eq. \eqref{eq:Eq05}, Eq. \eqref{eq:Eq04} is expressed as
\begin{equation}\label{eq:Eq06}
M \frac{\mathrm{d}v(t)}{\mathrm{d}t} + \6 \pi R \eta v(t) + \6 R^\2 \sqrt{\pi \rho_f \eta} \,\Gamma\left( \nicefrac{\1}{\2}\right) I^{\nicefrac{\1}{\2}} \left[\frac{\mathrm{d}v(t)}{\mathrm{d}t}  \right] = f_R (t)
\end{equation}
where $M=m+\frac{\1}{\2}m_f=\frac{\4}{\3}\pi R^\3 \left( \rho_p + \frac{\1}{\2}\rho_f \right)$ 
with $\rho_p=$ mass density of the Brownian particle. By using the properties of fractional calculus \citep{OldhamSpanier1974, SamkoKilbasMarichev1974, MillerRoss1993, Podlubny1998} 
the fractional integral of order $\nicefrac{\1}{\2}$ of the derivative of the velocity (first-order derivative) is the fractional derivative of order $\nicefrac{\1}{\2}$ of the velocity history
\begin{equation}\label{eq:Eq07}
I^{\nicefrac{\1}{\2}} \left[ \frac{\mathrm{d}v(t)}{\mathrm{d}t} \right] = \frac{\mathrm{d}^{\nicefrac{\1}{\2}} v(t)}{\mathrm{d}t^{\nicefrac{\1}{\2}}} = \frac{\1}{\Gamma\left( -\frac{\1}{\2}\right)} \int_{\0^-}^\1 \frac{v(\xi)}{(t-\xi)^{\nicefrac{\3}{\2}}} \, \mathrm{d}\xi
\end{equation}
More generally, the fractional derivative of order $q \in \mathbb{R}^+$ of a continuous function, $f(t)$, is defined within the context of generalized functions as the convolution of $f(t)$ with the kernel $\frac{\1}{\Gamma(-q)} \frac{\1}{t^{q+\1}}$, whereas, the Laplace transform of the fractional derivative of $f(t)$ is $\mathcal{L}\left\lbrace \frac{\mathrm{d}^q f(t)}{\mathrm{d}t^q} \right\rbrace=$ \break $\int_{\0^-}^\infty \frac{\mathrm{d}^q f(t)}{\mathrm{d}t^q} e^{-st}\, \mathrm{d}t=s^q \mathcal{L}\left\lbrace f(t) \right\rbrace=s^q\mathcal{F}(s)$ \citep{OldhamSpanier1974, MillerRoss1993, Mainardi2010, Makris2021a}. By employing the result of Eq. \eqref{eq:Eq07} in association with $\Gamma\left(\frac{\1}{\2}\right)=\sqrt{\pi}$, Eq. \eqref{eq:Eq06} simplifies to
\begin{equation}\label{eq:Eq08}
M\frac{\mathrm{d}v(t)}{\mathrm{d}t} + \6 \pi R^\2 \sqrt{\rho_f \eta} \frac{\mathrm{d}^{\nicefrac{\1}{\2}}v(t)}{\mathrm{d}t^{\nicefrac{\1}{\2}}}+\6 \pi R \eta v(t)=f_R(t)
\end{equation}
Equation \eqref{eq:Eq08} offers the remarkable result that fractional differentials do not only appear when modeling the phenomenological power-law relaxation of a wide range of viscoelastic materials \citep{Nutting1921, Gemant1936, ScottBlair1944, ScottBlair1947}, but also they emerge naturally from the solution of continuum mechanics equations as they result from fundamental conservation laws \citep{LandauLifshitz1959, ZwanzigBixon1970, Widom1971}.

The Langevin Eq. \eqref{eq:Eq04} or \eqref{eq:Eq08} which accounts for the hydrodynamic memory was solved analytically by Widom \cite{Widom1971} after solving an integral equation for the velocity autocorrelation function of the Brownian particle in association with the appropriate long-range correlation of the random process. Building on \citeauthor{Widom1971}'s (\citeyear{Widom1971}) solution the following result for the velocity autocorrelation function can be obtained \citep{ClercxSchram1992, LiRaizen2013}
\begin{align}\label{eq:Eq09}
\left\langle v(\0)v(t) \right\rangle = \frac{N K_B T}{M(b-a)} & \left[ b e^{b^\2 t} \text{Erfc}\left(b\sqrt{t}\right) \right. \\ \nonumber 
& \left. - ae^{a^\2 t} \text{Erfc}\left(a\sqrt{t}\right)  \right]
\end{align}
with
\begin{equation}\label{eq:Eq10}
a=\frac{z+\sqrt{z^\2-\4\zeta M}}{\2 M} \quad \text{and} \quad b=\frac{z-\sqrt{z^\2-\4\zeta M}}{\2 M}
\end{equation}
In this paper we follow the notation of Clercx and Schram \citep{ClercxSchram1992} where $z=\6\pi R^\2 \sqrt{\rho_f \eta}$ with units [M][L][T]$^{-\nicefrac{\3}{\2}}$ (say Pa s$^{\nicefrac{\3}{\2}}$) --- that is the coefficient of the $\nicefrac{\1}{\2}$ fractional derivative of the velocity of the Brownian particle appearing in the Langevin eq. \eqref{eq:Eq08}.

\section{The Dashpot--Inertpot--Inerter Parallel Connection}

For the analysis of classical Brownian motion, as was treated by \cite{UhlenbeckOrnstein1930} and \cite{WangUhlenbeck1945} (when effects of the hydrodynamic memory are neglected) of microparticles suspended in any linear, isotropic viscoelastic material, Makris \cite{Makris2020} presented a viscous--viscoelastic correspondence principle which reveals that the mean-square displacement and its time derivatives \citep{Makris2021SM} of the randomly moving Brownian microspheres with mass $m$ and radius $R$ are identical to the deterministic time response functions \bigg(creep compliance $J(t)=\frac{\3\pi R}{N K_B T}\left\langle \Delta r^\2 (t) \right\rangle$, impulse response function $h(t)=\frac{\3\pi R}{N K_B T}\frac{\mathrm{d}\left\langle \Delta r^\2 (t) \right\rangle}{\mathrm{d}t}$ and impulse strain--rate response function $\psi(t)=\frac{\3\pi R}{N K_B T}\frac{\mathrm{d}^\2\left\langle \Delta r^\2 (t) \right\rangle}{\mathrm{d}t^\2}=\frac{\6\pi R}{N K_B T}\left\langle v(\0)v(t) \right\rangle$\bigg) of a mechanical network that is a parallel connection of the linear viscoelastic material (within which the microspheres are immersed) with an inerter with distributed inertance $m_R=\frac{m}{\6\pi R}$.

\begin{figure}[t!]
\centering
\includegraphics[width=0.9\linewidth, angle=0]{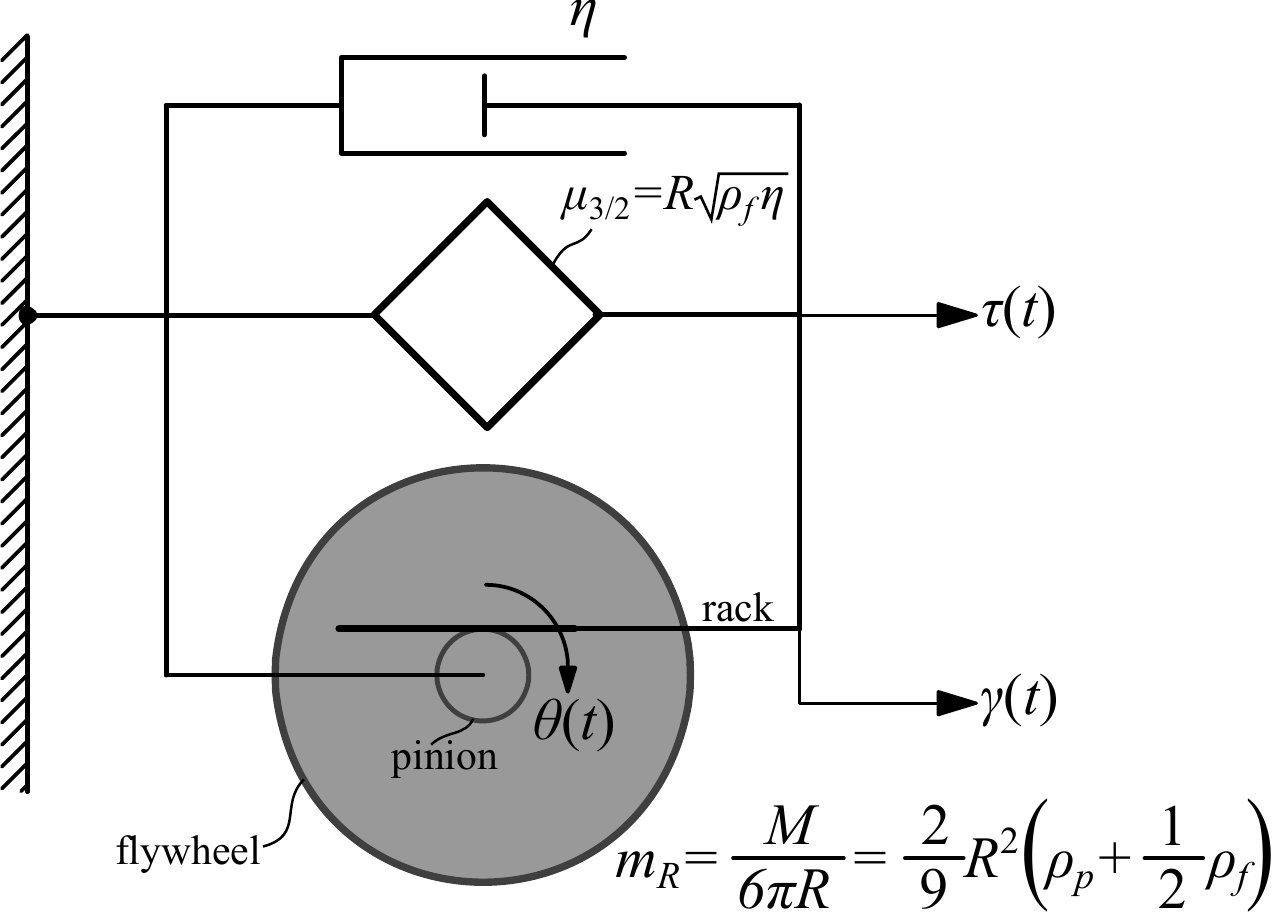}
\caption{The rheological analogue for Brownian motion in a Newtonian viscous fluid with hydrodynamic memory. A dashpot with viscosity $\eta$ is connected in parallel with a Scott-Blair element with material constant $\mu_{\nicefrac{\3}{\2}}=R\sqrt{\rho_f \eta}$ and an inerter with distributed inertance $m_R=\frac{M}{\6\pi R}=\frac{\2}{\text{9}}R^\2\rho_f\left( \frac{\rho_p}{\rho_f} + \frac{\1}{\2}\right)$ (the dashpot--inertpot--inerter parallel connection).}
\label{fig:Fig01}
\end{figure}

In this paper we build upon the viscous--viscoelastic correspondence principle for Brownian motion \citep{Makris2020, Makris2021SM} to develop a mechanical analogue for Brownian motion with hydrodynamic memory which simplifies appreciably the calculations of the velocity autocorrelation function and the mean-square displacement. The synthesis of the proposed macroscopic mechanical network is suggested from the left-hand side of the Langevin Eq. \eqref{eq:Eq08}, which consists of an inertia term, a $\nicefrac{\1}{\2}$-order fractional derivative of the velocity term and a viscous term acting in parallel to balance the random force $f_R(t)$. Accordingly, we examine the frequency and the time-response functions of the mechanical network shown in Fig. \ref{fig:Fig01} which is a parallel connection of a dashpot with shear viscosity $\eta$, a fractional Scott--Blair element with material constant $\mu_{\nicefrac{\3}{\2}}=R\sqrt{\rho_f\eta}=\frac{z}{\6\pi R}$ and an inerter with distributed inertance $m_R=\frac{M}{\6\pi R}=\frac{\2}{\text{9}}R^\2\rho_f\left(\frac{\rho_p}{\rho_f}+\frac{\1}{\2} \right)$.

Following Nutting's \cite{Nutting1921} observations that the stress response of several fluid-like materials when subjected to a step--strain $(\gamma(t)=U(t-\0))$ excitation decays following a power law: $\tau(t)=G_{\text{ve}}(t)\sim t^{-q}$ with 0 $\leqslant q \leqslant$ 1 and the early work of Gemant \cite{Gemant1936, Gemant1938} on fractional differentials, Scott-Blair \cite{ScottBlair1944, ScottBlair1947}, proposed the springpot element (0 $\leqslant q \leqslant$ 1) which is a mechanical idealization in--between an elastic spring and a viscous dashpot with constitutive law
\begin{equation}\label{eq:Eq11}
\tau(t)=\mu_q \frac{\mathrm{d}^q\gamma(t)}{\mathrm{d}t^q} \comma \quad q\in\mathbb{R}^+
\end{equation}
In this study we adopt the name `Scott--Blair element' compared to the restrictive `springpot' element since the order of differentiation of the fractional element appearing in Fig. \ref{fig:Fig01} is $q=\nicefrac{\3}{\2}$. In view that 1 $\leqslant q=\nicefrac{\3}{\2} \leqslant $ 2 and by adopting the terminology	 introduced by Scott--Blair (`springpot'), the fractional element appearing in Fig. \ref{fig:Fig01} is coined the `inertpot' since it is a mechanical idealization in--between an inerter ($q=$ 2) and a viscous dashpot ($q=$ 1).

An inerter is a linear mechanical element for which, at the force--displacement level, the output force is proportional only to the relative acceleration of its end-nodes (terminals) and its constant of proportionality is the inertance with units of mass [M] \citep{Smith2002, PapageorgiouSmith2005, Makris2017, Makris2018}. At the stress--strain level, the constant of proportionality of the inerter shown schematically in Fig. \ref{fig:Fig01} is the distributed inertance, $m_R$, with units [M][L]$^{-\1}$ (i.e Pa s$^\2$). Given the parallel connection of the dashpot--inertpot--inerter network shown in Fig. \ref{fig:Fig01}, its constitutive law is
\begin{equation}\label{eq:Eq12}
\tau(t)=\eta \frac{\mathrm{d}\gamma(t)}{\mathrm{d}t} + \mu_{\nicefrac{\3}{\2}} \frac{\mathrm{d}^{\nicefrac{\3}{\2}}\gamma(t)}{\mathrm{d}t^{\nicefrac{\3}{\2}}} + m_R \frac{\mathrm{d}^\2\gamma(t)}{\mathrm{d}t^\2}
\end{equation}
The Laplace transform of Eq. \eqref{eq:Eq12} gives
\begin{equation}\label{eq:Eq13}
\tau(s)=\mathcal{G}(s)\gamma(s)= \left( \eta s + \mu_{\nicefrac{\3}{\2}} s^{\nicefrac{\3}{\2}} + m_R s^\2 \right)\gamma(s)
\end{equation}
where $\mathcal{G}(s) = \left( \eta s + \mu_{\nicefrac{\3}{\2}} s^{\nicefrac{\3}{\2}} + m_R s^\2 \right)$ is the complex dynamic modulus. The complex dynamic compliance of the rheological network shown in Fig. \ref{fig:Fig01} is
\begin{align}\label{eq:Eq14}
\mathcal{J}(s) =\frac{\1}{\mathcal{G}(s)} & =\frac{\1}{\eta s + \mu_{\nicefrac{\3}{\2}} s^{\nicefrac{\3}{\2}} + m_R s^\2} \\ \nonumber 
& =\frac{\1}{m_R}\frac{\1}{s}\frac{\1}{\left(s+\frac{\mu_{\nicefrac{\3}{\2}} }{m_R} s^{\nicefrac{\1}{\2}} + \frac{\eta}{m_R}\right)}
\end{align}
The recently published viscous--viscoelastic correspondence principle for classical Brownian motion (without hydrodynamic memory) \citep{Makris2020} uncovered that ensemble averages of the motion of Brownian particles are proportional to macroscopic viscoelastic response functions. For instance the Laplace transform of the velocity autocorrelation function of Brownian particles satisfies the identity
\begin{equation}\label{eq:Eq15}
\left\langle v(s)v(\0)\right\rangle=\frac{N K_B T}{\6\pi R} s \mathcal{J}(s)
\end{equation}
where $s\mathcal{J}(s) = \phi(s)=\frac{\dt{\gamma}(s)}{\tau(s)}$ is the complex dynamic fluidity (admittance) of the rheological network that resists the random forces, $f_R(t)$. In this paper we extend this concept for the case with hydrodynamic memory and is implemented to the Langevin Eq. \eqref{eq:Eq08}. Accordingly, substitution of Eq. \eqref{eq:Eq14} into Eq. \eqref{eq:Eq15} gives
\begin{align}\label{eq:Eq16}
\left\langle v(s)v(\0)\right\rangle & = \frac{N K_B T}{\6\pi R} \frac{\1}{m_R} \frac{\1}{s+\frac{\mu_{\nicefrac{\3}{\2}} }{m_R} s^{\nicefrac{\1}{\2}} + \frac{\eta}{m_R}} \\ \nonumber 
& = \frac{N K_B T}{M} \frac{\1}{s+\frac{z}{M} s^{\nicefrac{\1}{\2}} + \frac{\zeta}{M}}
\end{align}
where $m_R=\frac{M}{\6\pi R}=\frac{\2}{\text{9}}R^\2\rho_f \left( \frac{\rho_p}{\rho_f}+\frac{\1}{\2} \right) $,  $\mu_{\nicefrac{\3}{\2}}=R\sqrt{\rho_f\eta}=\frac{z}{\6\pi R}$ and the last expression of Eq. \eqref{eq:Eq16} follows the Clercx and Schram \cite{ClercxSchram1992} notation. The inverse Laplace transform of Eq. \eqref{eq:Eq16} is evaluated by employing the fraction expansion
\begin{equation}\label{eq:Eq17}
\frac{\1}{s+(a+b)s^{\nicefrac{\1}{\2}}+ab}=\frac{\1}{b-a}\left( \frac{\1}{s^{\nicefrac{\1}{\2}}+a} - \frac{\1}{s^{\nicefrac{\1}{\2}}+b} \right)
\end{equation}
together with the result \citep{Erdelyi1954}
\begin{align}\label{eq:Eq18}
\mathcal{L}^{-\1}\left\lbrace \frac{\1}{s^{\nicefrac{\1}{\2}}+a} \right\rbrace & = \frac{\1}{\sqrt{\pi t}} -a e^{a^\2 t} \text{Erfc}(a\sqrt{t}) \\ \nonumber 
&=\frac{\1}{\sqrt{t}}E_{\nicefrac{\1}{\2},\nicefrac{\1}{\2}}(-a\sqrt{t})
\end{align}
where Erfc$(x)$ is the complementary error function and $E_{p, q} (x)$ is the two parameter Mittag--Leffler function \citep{Erdelyi1953, GorenfloKilbasMainardiRogosin2014}
\begin{equation}\label{eq:Eq19newA}
E_{p, q}(x)= \sum_{j=\0}^{\infty} \frac{x^j}{\Gamma(jp+q)}\comma \quad p, q > \0
\end{equation}

With reference to the denominators of the left-hand side of Eqs. \eqref{eq:Eq17} and the right-hand side of Eq. \eqref{eq:Eq16}, the solution of the system of equations $a+b=\frac{z}{M}$ and $ab=\frac{\zeta}{M}$ yields the values of $a$ and $b$ offered by Eq. \eqref{eq:Eq10}.

By virtue of Eqs. \eqref{eq:Eq17} and \eqref{eq:Eq18}, the inverse Laplace transform of Eq. \eqref{eq:Eq16} yields the result offered by Eq. \eqref{eq:Eq09}. Consequently, the analysis presented herein by implementing the viscous--viscoelastic correspondence principle \citep{Makris2020, Makris2021SM} confirms that the velocity autocorrelation function of Brownian particles suspended in a viscous Newtonian fluid where effects of hydrodynamic memory are accounted is given by:
\begin{equation}\label{eq:Eq19}
\left\langle v(\0)v(t)\right\rangle = \frac{N K_B T}{\6\pi R} \psi(t)
\end{equation}
where $\psi(t) = \mathcal{L}^{-\1}\left\lbrace \phi(s) \right\rbrace = \mathcal{L}^{-\1}\left\lbrace s \mathcal{J}(s) \right\rbrace$ is the impulse strain-rate response function of the mechanical network shown in Fig. \ref{fig:Fig01} (the dashpot--inertpot--inerter parallel connection), defined as the resulting strain-rate output, $\dt{\gamma}(t)$, due to an impulsive stress input, $\tau(t) = \delta(t-\0)$.

In terms of the two-parameter Mittag--Leffler function given by the identity \eqref{eq:Eq18}, an alternative expression to Eq. \eqref{eq:Eq09} for the velocity autocorrelation function of Brownian particles immersed in a viscous, Newtonian fluid with hydrodynamic memory is
\begin{align}\label{eq:Eq44new}
\left\langle v(\0)v(t) \right\rangle & = \frac{\1}{\2} \frac{\mathrm{d}^\2\left\langle \Delta r^\2 (t) \right\rangle}{\mathrm{d}t^\2} = \frac{N K_B T}{M (b-a)} \frac{\1}{\sqrt{t}} \\ \nonumber
& \times \left[ E_{\nicefrac{\1}{\2}, \nicefrac{\1}{\2}} \left( -a\sqrt{t} \right) - E_{\nicefrac{\1}{\2}, \nicefrac{\1}{\2}} \left( -b\sqrt{t} \right) \right]
\end{align}

The constants $a$ and $b$ offered by Eq. \eqref{eq:Eq10} involve the radical
\begin{align}\label{eq:Eq20}
\sqrt{z^\2-\4\zeta M} & = z \sqrt{\1-\frac{\text{8}}{\text{9}} \left( \frac{1}{2} + \frac{\rho_p}{\rho_f} \right)} \\ \nonumber 
& = \6 \pi R^\2 \sqrt{\rho_f \eta} \sqrt{\1-\frac{\text{8}}{\text{9}} \left( \frac{1}{2} + \frac{\rho_p}{\rho_f} \right)}
\end{align}
which remains real as long as $\rho_p < \frac{5}{8} \rho_f$. For most practical cases $\rho_p > \rho_f$. As an example, when melanine rasin microspheres ($\rho_p = \text{1570}$ kg/m$^\3$) are immensed in water ($\rho_f = \text{1000}$ kg/m$^\3$), $\rho_p = \text{1.57} \rho_f$ \citep{Franosch_etal2011, DominguezGarciaCardinauxBertsevaForroScheffoldJeney2014}; whereas, when they are immensed in acetone ($\rho_f = \text{784}$ kg/m$^\3$), $\rho_p = \2 \rho_f$. Accordingly, for most practical cases, the radical $\sqrt{z^2 - \text{4}\zeta M}$ is an imaginary quantity
\begin{equation}\label{eq:Eq21}
\sqrt{z^\2-\text{4} \zeta M} = \operatorname{i} z \sqrt{\frac{\text{8}}{\text{9}} \left( \frac{1}{2} + \frac{\rho_p}{\rho_f} \right)-\1} = \operatorname{i} z \sqrt{\sigma-\1}
\end{equation}
where $\operatorname{i}=\sqrt{-\1}$ and
\begin{equation}\label{eq:Eq22}
\sigma = \frac{\text{8}}{\text{9}} \left( \frac{\rho_p}{\rho_f} + \frac{\1}{\2} \right)
\end{equation}
is a parameter of the Brownian particle--Newtonian fluid system described by the Langevin Eq. \eqref{eq:Eq04} of \eqref{eq:Eq08}. With the introduction of the parameter $\sigma$ defined by Eq. \eqref{eq:Eq22}, the constants $a$ and $b$ introduced by \cite{ClercxSchram1992} and offered by Eq. \eqref{eq:Eq10} are expressed as $a=\frac{z}{\2 M} (\1+\operatorname{i}\sqrt{\sigma-\1})$ and $b=\frac{z}{\2 M} (\1-\operatorname{i}\sqrt{\sigma-\1})$, whereas, the factor
\begin{equation}\label{eq:Eq23}
\frac{z}{\2 M} = \frac{\text{9}}{\text{4}} \frac{\1}{R} \frac{\sqrt{\eta}}{\sqrt{\rho_f}} \frac{\1}{\left( \frac{\rho_p}{\rho_f} + \frac{\1}{\2} \right)} = \frac{\2}{\sigma \sqrt{\tau_f}}
\end{equation}
where $\sigma$ is defined by Eq. \eqref{eq:Eq22} and $\tau_f = \frac{R^\2 \rho_f}{\eta}$ is a time-scale which describes the time needed by the perturbed fluid flow to diffuse over one Brownian microsphere radius \citep{PaulPusey1981, Franosch_etal2011, JannaschMahamdehSchaffer2011}. Accordingly, for $\sigma > \1 \left( \frac{\rho_p}{\rho_f} > \frac{\text{5}}{\text{8}} \right)$ the constants $a$ and $b$ introduced by Eq. \eqref{eq:Eq10} assume the expressions
\begin{equation}\label{eq:Eq24}
a=\frac{\2}{\sigma \sqrt{\tau_f}} \left( \1+\operatorname{i}\sqrt{\sigma-\1} \right) \enskip \text{and} \enskip b=\frac{\2}{\sigma \sqrt{\tau_f}} \left( \1-\operatorname{i}\sqrt{\sigma-\1} \right)
\end{equation}
and the expression for the velocity autocorrelation function offered by Eq. \eqref{eq:Eq09} gives
\begin{figure}[b!]
\centering
\includegraphics[width=\linewidth, angle=0]{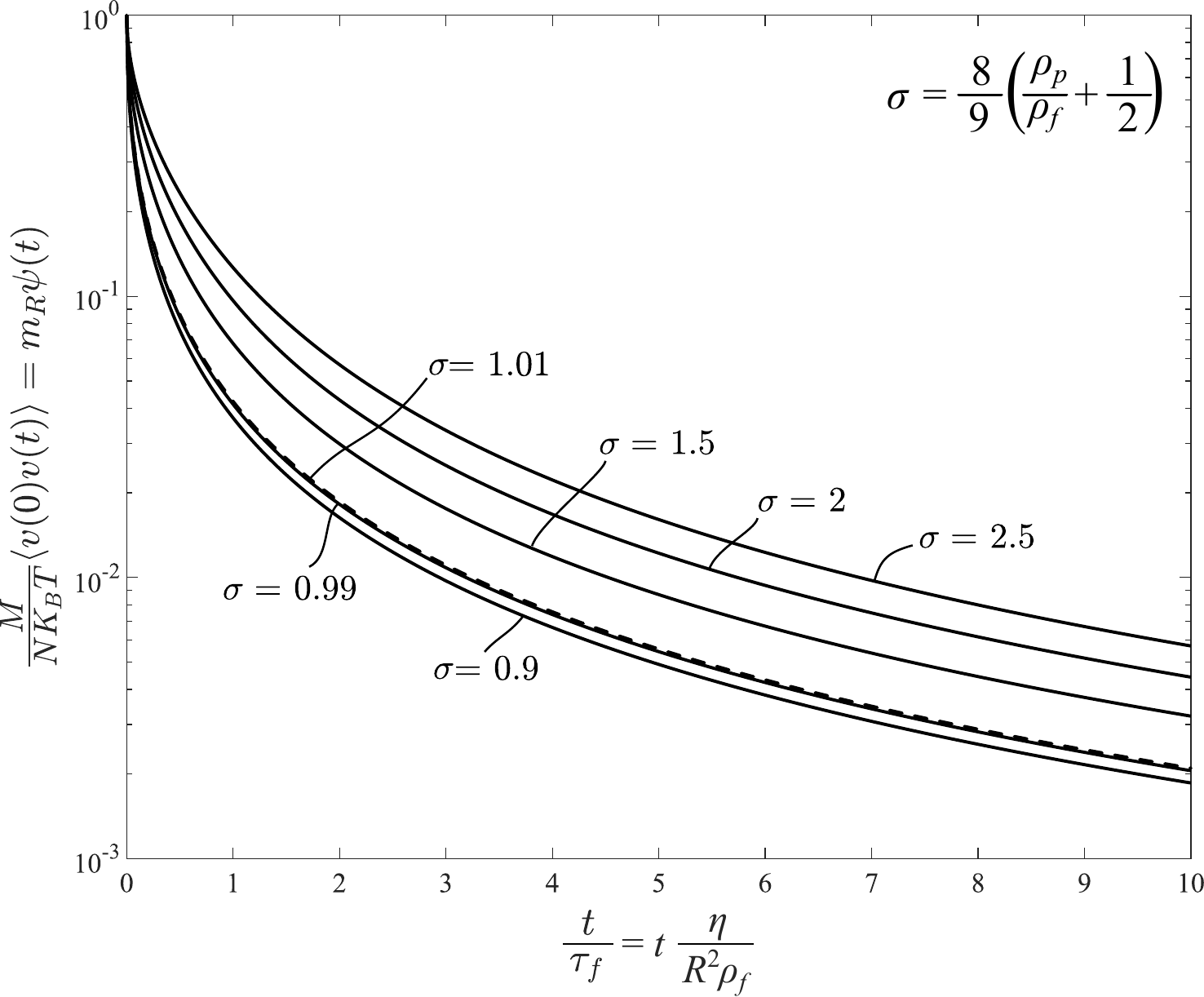}
\caption{Normalized velocity autocorrelation function of Brownian particles immersed in a Newtonian, viscous fluid with viscosity $\eta$; when effects from hydrodynamic memory are accounted. The curves for $\sigma = \frac{\text{8}}{\text{9}}\left( \frac{\rho_p}{\rho_f} + \frac{\1}{\2} \right)=$ 0.9 and 0.99 are computed with Eq. \eqref{eq:Eq27}, whereas the curves for $\sigma=$ 1.01, 1.5, 2 and 2.5 are computed with Eq. \eqref{eq:Eq25} showing the continuity of the solution as $\sigma$ crosses the value of $\sigma=$ 1.}
\label{fig:Fig02}
\end{figure}
\begin{widetext}
\begin{align}\label{eq:Eq25}
\left\langle v(\0)v(t)\right\rangle & = \frac{N K_B T}{M} \frac{e^{-\frac{\4(\sigma-\2)}{\sigma^\2} \frac{t}{\tau_f}}}{\2\sqrt{\sigma-\1}} \Bigg[ \left( \sqrt{\sigma-\1}+\operatorname{i} \right) e^{-\frac{\operatorname{i}\text{8}\sqrt{\sigma-\1}}{\sigma^\2} \frac{t}{\tau_f}} \text{Erfc} \left( \frac{\2 \left( \1-\operatorname{i}\sqrt{\sigma-\1} \right)}{\sigma} \sqrt{\frac{t}{\tau_f}} \right) \\ \nonumber
& + \left( \sqrt{\sigma-\1}-\operatorname{i} \right) e^{+\frac{\operatorname{i}\text{8}\sqrt{\sigma-\1}}{\sigma^\2} \frac{t}{\tau_f}} \text{Erfc} \left( \frac{\2 \left( \1+\operatorname{i}\sqrt{\sigma-\1} \right)}{\sigma} \sqrt{\frac{t}{\tau_f}} \right) \Bigg]\comma \quad \sigma > \1
\end{align}
\end{widetext}
The velocity autocorrelation function offered by Eq. \eqref{eq:Eq25} for $\sigma>\1$ is a real-valued function given that the imaginary parts of the right-hand side of Eq. \eqref{eq:Eq25} cancel. For the case when $\sigma < \1 \left( \frac{\rho_p}{\rho_f} < \frac{\text{5}}{\text{8}} \right)$, the parameters $a$ and $b$ introduced by \cite{ClercxSchram1992} assume the expressions
\begin{equation}\label{eq:Eq26}
a=\frac{\2}{\sigma \sqrt{\tau_f}} \left( \1+\sqrt{\1-\sigma} \right) \quad \text{and} \quad b=\frac{\2}{\sigma \sqrt{\tau_f}} \left( \1-\sqrt{\1-\sigma} \right)
\end{equation}
and the expession for the velocity autocorrelation function given by Eq. \eqref{eq:Eq09} yields
\begin{widetext}
\begin{align}\label{eq:Eq27}
\left\langle v(\0)v(t)\right\rangle & = \frac{N K_B T}{M} \frac{e^{\frac{\4(\2-\sigma)}{\sigma^\2} \frac{t}{\tau_f}}}{\2\sqrt{\1-\sigma}} \Bigg[ -\left( \1-\sqrt{\1-\sigma} \right) e^{-\frac{\text{8}\sqrt{\1-\sigma}}{\sigma^\2} \frac{t}{\tau_f}} \text{Erfc} \left( \frac{\2 \left( \1-\sqrt{\1-\sigma} \right)}{\sigma} \sqrt{\frac{t}{\tau_f}} \right) \\ \nonumber
& + \left( \1+ \sqrt{\1-\sigma} \right) e^{+\frac{\text{8}\sqrt{\1-\sigma}}{\sigma^\2} \frac{t}{\tau_f}} \text{Erfc} \left( \frac{\2 \left( \1+\sqrt{\1+\sigma} \right)}{\sigma} \sqrt{\frac{t}{\tau_f}} \right) \Bigg] \comma \quad \sigma < \1
\end{align}
\end{widetext}
Figure \ref{fig:Fig02} plots the normalized velocity autocorrelation function $\frac{M}{N K_B T} \left\langle v(\0)v(t) \right\rangle = m_R \psi(t)$ for various values of the parameter $\sigma$. For $\sigma=\text{0.9}$ and 0.99 the cuves are computed with Eq. \eqref{eq:Eq27}; whereas, for values of $\sigma=\text{1.01}$, 1.5, 2 and 2.5 the curves are computed with Eq. \eqref{eq:Eq25}.

\section{Time Derivative of the Mean-Square Displacement of Brownian Particles in a Viscous Fluid with Hydrodynamic Memory}

The time derivative of the mean square displacement of Brownian particles has been occasionally interpreted as a time-dependent diffusion coefficient $\frac{\mathrm{d}\left\langle \Delta r^\2 (t) \right\rangle}{\mathrm{d}t}=\2 ND(t)$ \citep{PaulPusey1981,WeitzPinePuseyTough1989,SegrePusey1996,Sperl2005,SafdariCherstvyChechkinBodrovaMetzler2017} and has a central role in the interpretation of Brownian motion at various time scales \citep{KhanMason2014} and confined spacings \citep{GhoshKrishnamurthy2018}.

With reference to Eq. \eqref{eq:Eq15} in association that the velocity autocorrelation function $\left\langle v(\0)v(t) \right\rangle=\frac{\1}{\2}\frac{\mathrm{d}^\2 \left\langle \Delta r^\2 (t) \right\rangle}{\mathrm{d}t^\2}$ is the second time-derivative of the mean-square displacement \citep{KenkreKuhneReineker1981, BianKimKarniadakis2016}; in a recent paper Makris \cite{Makris2021SM} uncovered that the complex dynamic compliance, $\mathcal{J}(s)$, of the rheological analogue for Brownian motion, such as the one shown in Fig. \ref{fig:Fig01} is proportional to the Laplace transform of the time-derivative of the mean-square displacement
\begin{equation}\label{eq:Eq28}
\mathcal{J}(s)=\frac{\3\pi R}{N K_B T} \mathcal{L}\left\lbrace \frac{\mathrm{d} \left\langle \Delta r^\2 (t) \right\rangle}{\mathrm{d}t} \right\rbrace
\end{equation}

The inverse Laplace transform of the complex dynamic compliance, $\mathcal{J}(s)$ given by Eq. \eqref{eq:Eq14} is the impulse fluidity, $\phi(t)=\mathcal{L}^{-\1}\left\lbrace \mathcal{J}(s)\right\rbrace$ \citep{Giesekus1995, MakrisKampas2009, MakrisEfthymiou2020}, defined as the resulting strain $\gamma(t)$, due to an impulse stress input, $\tau(t)=\delta(t-\0)$. The impulse function $\delta(t-\0)$ is the Dirac delta function \citep{Lighthill1958} with the property $\mathcal{L}\left\lbrace \delta(t-\xi)\right\rbrace=\int_{\0^-}^\infty \delta(t-\xi)e^{-st} \, \mathrm{d}t = e^{-\xi s}$. The equivalent of the impulse fluidity, $\phi(t)$, at the displacement--force level is the impulse response function often expressed with $h(t)$. Given that the term ``impulse response function'' (rather than the term ``impulse fluidity'') is widely known and used in dynamics \citep{CloughPenzien1970}, structural mechanics \citep{HarrisCrede1976}, electrical signal processing \citep{OppenheimSchafer1975, Reid1983} and economics \citep{BorovivckaHansen2016}, Makris \cite{Makris2021SM} adopted the term ``impulse response function $=h(t)$'', rather than the term impulse fluidity'' used narrowly in the viscoelasticity literature alone. Accordingly, from Eq. \eqref{eq:Eq14} the impulse response function of the dashpot--interpot--inerter parallel connection is
\begin{equation}\label{eq:Eq29}
h(t)=\mathcal{L}^{-\1}\left\lbrace \mathcal{J}(s) \right\rbrace=\frac{\1}{m_R} \mathcal{L}^{-\1}\left\lbrace \frac{\1}{s} \frac{\1}{\left(s+\frac{\mu_{\nicefrac{\3}{\2}}}{m_R}s^{\nicefrac{\1}{\2}}+\frac{\eta}{m_R}\right)} \right\rbrace
\end{equation}
where again $m_R=\frac{M}{\6 \pi R}=\frac{\2}{\text{9}}R^\2\rho_f\left( \frac{\rho_p}{\rho_f}+\frac{\1}{\2} \right)$ and $\mu_{\nicefrac{\3}{\2}}=R\sqrt{\rho_f \eta}=\frac{z}{\6 \pi R}$ with $z$ being the Clercx and Schram \cite{ClercxSchram1992} parameter appearing in Eq. \eqref{eq:Eq10}. By following the Clercx and Schram \cite{ClercxSchram1992} notation in association with Eq. \eqref{eq:Eq28}, the impulse response function given by Eq. \eqref{eq:Eq29} is
\begin{align}\label{eq:Eq30}
h(t)&=\frac{\3 \pi R}{N K_B T} \frac{\mathrm{d}\left\langle \Delta r^\2 (t) \right\rangle}{\mathrm{d}t} \\ \nonumber
&= \frac{\6 \pi R}{M} \mathcal{L}^{-\1} \left\lbrace \frac{\1}{s} \frac{\1}{\left(s+\frac{z}{M}s^{\nicefrac{\1}{\2}}+\frac{\zeta}{M}\right)} \right\rbrace
\end{align}
The inverse Laplace transform of Eq. \eqref{eq:Eq29} is evaluated by employing the fraction expansion
\begin{align}\label{eq:Eq31}
& \frac{\1}{s \left[ s+(a+b)s^{\nicefrac{\1}{\2}}+ab \right]}\\ \nonumber 
& =\frac{\1}{b-a} \left[ \frac{\1}{s(s^{\nicefrac{\1}{\2}}+a)} -\frac{\1}{s(s^{\nicefrac{\1}{\2}}+b)} \right]
\end{align}
together with the result \citep{Erdelyi1954}
\begin{align}\label{eq:Eq32}
\mathcal{L}^{-\1} \left\lbrace \frac{\1}{s(s^{\nicefrac{\1}{\2}}+a)} \right\rbrace &=\frac{\1}{a} \left[ \1-e^{a^\2t} \text{Erfc}(a\sqrt{t}) \right]\\ \nonumber
&=\sqrt{t}E_{\nicefrac{\1}{\2},\nicefrac{\3}{\2}}(-a\sqrt{t})
\end{align}
With reference to the denominators of the left-hand side of Eq. \eqref{eq:Eq31} and the right-hand side of Eq. \eqref{eq:Eq30}, the solution of the system of equations $a+b=\frac{z}{M}$ and $ab=\frac{\zeta}{M}$ yields the values of $a$ and $b$ offered by Eq. \eqref{eq:Eq10}. Accordingly, by virtue of Eqs. \eqref{eq:Eq31} and \eqref{eq:Eq32}, the impulse response function for Brownian motion in a Newtonian viscous fluid with hydrodynamic memory is
\begin{widetext}
\begin{equation}\label{eq:Eq33}
h(t)=\frac{\3 \pi R}{N K_B T} \frac{\mathrm{d}\left\langle \Delta r^\2 (t) \right\rangle}{\mathrm{d}t}= \frac{\6 \pi R}{M} \frac{\1}{b-a} \left\lbrace \frac{\1}{a} \left[ \1-e^{a^\2t} \text{Erfc}(a\sqrt{t}) \right] - \frac{\1}{b} \left[ \1-e^{b^\2t} \text{Erfc}(b\sqrt{t}) \right] \right\rbrace
\end{equation}
\end{widetext}
When $\frac{\rho_p}{\rho_f}>\frac{\text{5}}{\text{8}}$, parameter $\sigma>\1$ and the parameters $a$ and $b$ appearing in Eq. \eqref{eq:Eq33} are complex-valued and are offered by Eq. \eqref{eq:Eq24}. In this case ($\sigma>\1$) Eq. \eqref{eq:Eq33} becomes
\begin{widetext}
\begin{align}\label{eq:Eq34}
h(t)  =\frac{\3\pi R}{N K_B T}\frac{\mathrm{d}\left\langle \Delta r^\2 (t) \right\rangle}{\mathrm{d}t} & =\frac{\1}{\eta}\frac{\sigma}{\2\sqrt{\sigma-\1}} \Bigg\lbrace \frac{\operatorname{i}}{\1+\operatorname{i}\sqrt{\sigma-\1}} \Bigg[\1-e^{-\frac{\text{4}(\sigma-\2)}{\sigma^\2}\frac{t}{\tau_f}} e^{\frac{\operatorname{i}\text{8}\sqrt{\sigma-\1}}{\sigma^\2}\frac{t}{\tau_f}} \text{Erfc}\left( \frac{\2\left( \1+\operatorname{i}\sqrt{\sigma-\1} \right)}{\sigma} \sqrt{\frac{t}{\tau_f}}\right)\Bigg] \\ \nonumber
&  -\frac{\operatorname{i}}{\1-\operatorname{i}\sqrt{\sigma-\1}} \Bigg[\1-e^{-\frac{\text{4}(\sigma-\2)}{\sigma^\2}\frac{t}{\tau_f}} e^{\frac{\operatorname{i}\text{8}\sqrt{\sigma-\1}}{\sigma^\2}\frac{t}{\tau_f}} \text{Erfc}\left( \frac{\2\left( \1-\operatorname{i}\sqrt{\sigma-\1} \right)}{\sigma} \sqrt{\frac{t}{\tau_f}}\right)\Bigg]  \Bigg\rbrace \comma \quad \sigma>\1
\end{align}
\end{widetext}
The impulse response function, $h(t)$, offered by Eq. \eqref{eq:Eq34} for $\sigma>\1$ is a real-valued function given that the imaginary parts of the right-hand side of Eq. \eqref{eq:Eq34} cancel.

\begin{figure}[t!]
\centering
\includegraphics[width=\linewidth, angle=0]{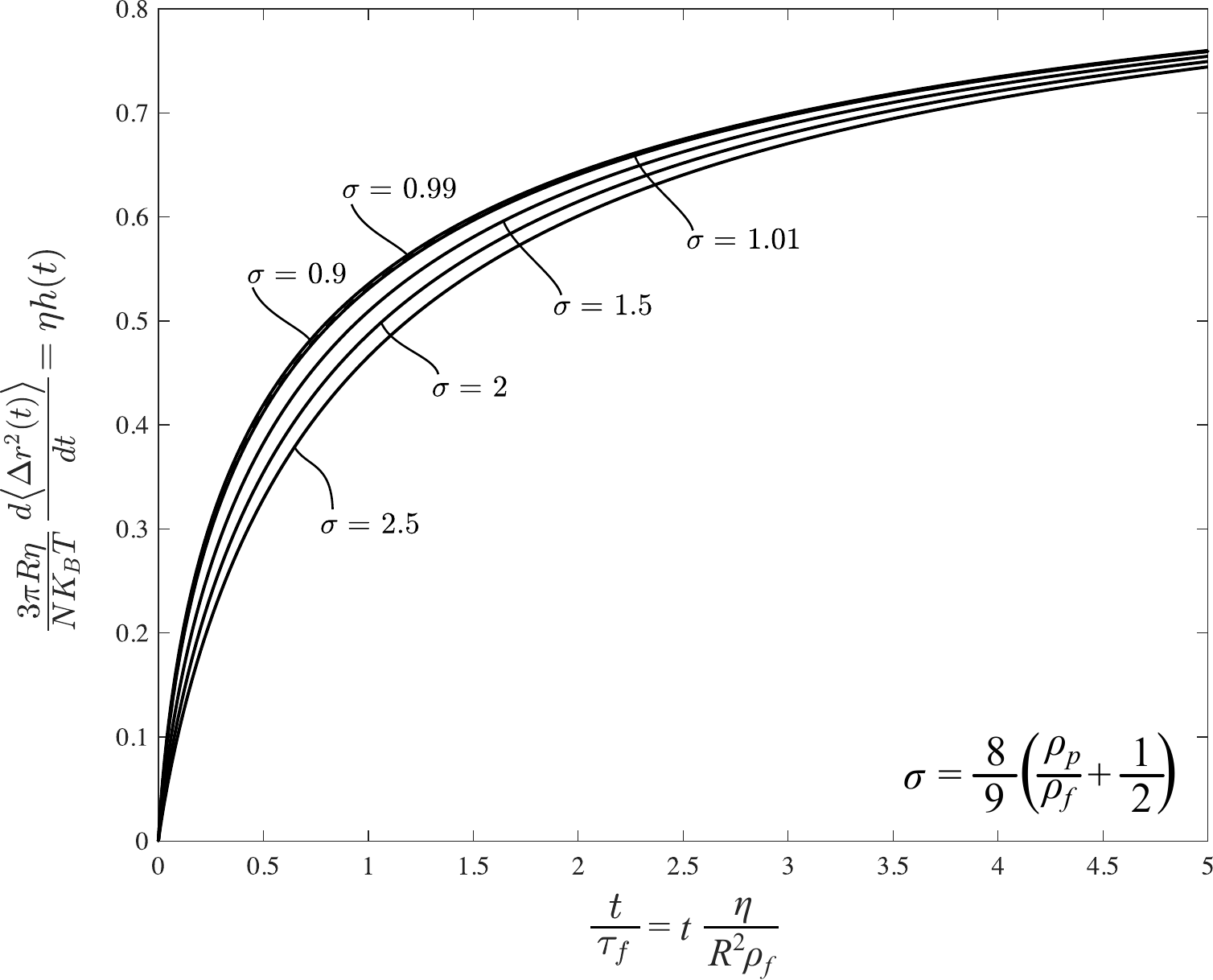}
\caption{Normalized time-derivative of the mean-square displacement of Brownian particles immersed in a Newtonian, viscous fluid with viscosity $\eta$; when effects from hydrodynamic memory are accounted. The curves for $\sigma=\frac{\text{8}}{\text{9}}\left( \frac{\rho_p}{\rho_f} +\frac{\1}{\2}\right)=$ 0.9 and 0.99 are computed with Eq. \eqref{eq:Eq35}; whereas, the curves for $\sigma=$ 1.01, 1.5, 2 and 2.5 are computed with Eq. \eqref{eq:Eq34} showing the continuity of the solution as parameter $\sigma$ crosses the value of $\sigma=\1$.}
\label{fig:Fig03}
\end{figure}

For the case when $\sigma<\1$ $\left(\frac{\rho_p}{\rho_f}<\frac{\text{5}}{\text{8}}\right)$, the parameters $a$ and $b$ appearing in Eq. \eqref{eq:Eq33} are real-valued and are offered by Eq. \eqref{eq:Eq26}. In this case ($\sigma<\1$), Eq. \eqref{eq:Eq33} becomes
\begin{widetext}
\begin{align}\label{eq:Eq35}
h(t) =\frac{\3\pi R}{N K_B T}\frac{\mathrm{d}\left\langle \Delta r^\2 (t) \right\rangle}{\mathrm{d}t} & =\frac{\1}{\eta}\frac{\sigma}{\2\sqrt{\1-\sigma}} \Bigg\lbrace -\frac{\1}{\1+\sqrt{\1-\sigma}} \Bigg[\1-e^{\frac{\text{4}}{\sigma^\2}\left(\2-\sigma+\2\sqrt{\1-\sigma}\right)\frac{t}{\tau_f}} \text{Erfc}\left( \frac{\2\left( \1+\sqrt{\1-\sigma} \right)}{\sigma} \sqrt{\frac{t}{\tau_f}}\right)\Bigg] \\ \nonumber
& +\frac{\1}{\1-\sqrt{\1-\sigma}} \Bigg[\1-e^{\frac{\text{4}}{\sigma^\2}\left(\2-\sigma-\2\sqrt{\1-\sigma}\right)\frac{t}{\tau_f}} \text{Erfc}\left( \frac{\2\left( \1-\sqrt{\1-\sigma} \right)}{\sigma} \sqrt{\frac{t}{\tau_f}}\right)\Bigg]  \Bigg\rbrace \comma \quad \sigma<\1
\end{align}
\end{widetext}
Figure \ref{fig:Fig03} plots the normalized time-derivative of the mean-square displacement $\frac{\3\pi R\eta}{N K_B T}\frac{\mathrm{d}\left\langle \Delta r^\2 (t) \right\rangle}{\mathrm{d}t}=\eta h(t)$ of Brownian particles with mass $m$ and radius $R$ immersed in a Newtonian, viscous fluid with viscosity $\eta$ for various values of the parameter $\sigma$ which accounts for hydrodynamic memory. The curves for $\sigma=$ 0.9 and 0.99 are computed with Eq. \eqref{eq:Eq35}; whereas the curves for $\sigma=$ 1.01, 1.5, 2 and 2.5 are computed with Eq. \eqref{eq:Eq34} and follow qualitatively the trend of the experimentally measured time-derivative of the mean-square displacement reported by \cite{PaulPusey1981} and \cite{WeitzPinePuseyTough1989}.

In terms of the two-parameter Mittag--Leffler function given by the identity \eqref{eq:Eq32}, an alternative expression to Eq. \eqref{eq:Eq33} for the time-derivative of the mean-square displacement of Brownian particles in a Newtonian fluid with hydrodynamic memory is:
\begin{align}\label{eq:Eq39new}
\frac{\mathrm{d}\left\langle \Delta r^\2 (t) \right\rangle}{\mathrm{d}t} & = \frac{N K_B T}{\3 \pi R} h(t) = \frac{\text{2} N K_B T}{M} \frac{\sqrt{t}}{b-a} \\ \nonumber
& \times \left[ E_{\nicefrac{\1}{\2}, \nicefrac{\3}{\2}} \left( -a\sqrt{t} \right) - E_{\nicefrac{\1}{\2}, \nicefrac{\3}{\2}} \left( -b\sqrt{t} \right) \right]
\end{align}
where again $a$ and $b$ are given by Eq. \eqref{eq:Eq10}.

\section{Mean-Square Displacement of Brownian Particles in a Newtonian Viscous Fluid with Hydrodynamic Memory}

According to the properties of the Laplace transform of the derivative of a function
\begin{equation}\label{eq:Eq42}
\mathcal{L}\left\lbrace \frac{\mathrm{d}\left\langle \Delta r^\2(t) \right\rangle}{\mathrm{d}t} \right\rbrace = s \left\langle \Delta r^\2(s) \right\rangle-\left\langle \Delta r^\2(\0) \right\rangle
\end{equation}
Given that at the time origin $(t=\0)$, $\left\langle \Delta r^\2(\0) \right\rangle=\0$, substitution of Eq. \eqref{eq:Eq42} into Eq. \eqref{eq:Eq28} gives
\begin{equation}\label{eq:Eq43}
\left\langle \Delta r^\2(s) \right\rangle = \frac{N K_B T}{\3\pi R} \frac{\mathcal{J}(s)}{s}=\frac{N K_B T}{\3\pi R} \mathcal{L}\left\lbrace J(t) \right\rbrace
\end{equation}
where $\mathcal{J}(s)$ is the complex dynamic compliance of the rheological network shown in Fig. \ref{fig:Fig01} and expressed by Eq. \eqref{eq:Eq14}; whereas, $J(t)$ is the creep compliance of the rheological network in the time-domain. Accordingly from Eq. \eqref{eq:Eq43} in association with Eq. \eqref{eq:Eq14}
\begin{align}\label{eq:Eq44}
\mathcal{L}\left\lbrace J(t) \right\rbrace & =\frac{\1}{m_R}\frac{\1}{s^\2}\frac{\1}{\left( s+\frac{\mu_{\3/\2}}{m_R}s^{\1/\2}+\frac{\eta}{m_R} \right)} \\ \nonumber
& =\frac{\6\pi R}{M}\frac{\1}{s^\2}\frac{\1}{\left( s+\frac{z}{M}s^{\1/\2}+\frac{\zeta}{M} \right)}
\end{align}
The inverse Laplace transform of Eq. \eqref{eq:Eq44} is evaluated by employing the fraction expansion
\begin{align}\label{eq:Eq45}
& \frac{\1}{s^\2\left( s+(a+b)s^{\1/\2}+ab \right)} \\ \nonumber 
& =\frac{\1}{b-a}\left[ \frac{\1}{s^\2\left( s^{\1/\2} + a \right)} - \frac{\1}{s^\2\left( s^{\1/\2} + b \right)} \right]
\end{align}
together with the known result \citep{GorenfloKilbasMainardiRogosin2014}
\begin{equation}\label{eq:Eq36}
\mathcal{L}^{-\1} \left\lbrace \frac{s^{p-q}}{s^p+a} \right\rbrace = t^{q-\1} E_{p, q}(-at^p) \comma \quad p < q\in \mathbb{R}.
\end{equation}
Upon taking $p=\frac{\1}{\2}$ and $q=\frac{\text{5}}{\2}$, the Laplace transform of any of the two fractions appearing on the right-hand side of Eq. \eqref{eq:Eq45} is
\begin{align}\label{eq:Eq46}
\mathcal{L}^{-\1}\left\lbrace \frac{s^{\1/\2-\text{5}/\2}}{s^{\1/\2}+a} \right\rbrace & =\mathcal{L}^{-\1}\left\lbrace \frac{\1}{s^\2\left(s^{\1/\2}+a\right)} \right\rbrace \\ \nonumber
& = t^{\3/\2}E_{\1/\2\comma\,\text{5}/\2}(-at^{\1/\2})
\end{align}
With reference to the denominators of the left-hand side of Eq. \eqref{eq:Eq45} and the right-hand side of Eq. \eqref{eq:Eq44}, the solution of the system of equations $a+b=\frac{z}{M}$ and $ab=\frac{\zeta}{M}$ yields the values of $a$ and $b$ offered by Eq. \eqref{eq:Eq10}. Accordingly, by virtue of Eqs. \eqref{eq:Eq45} and \eqref{eq:Eq46}, the mean-square displacement of Brownian particles in a Newtonian viscous fluid with hydrodynamic memory is
\begin{align}\label{eq:Eq47}
\left\langle \Delta r^\2(t) \right\rangle & = \frac{N K_B T}{\3\pi R} J(t)= \frac{\2 N K_B T}{M} \frac{t^{\3/\2}}{b-a} \\ \nonumber
& \times \left[ E_{\1/\2\comma\,\text{5}/\2}(-a\sqrt{t}) - E_{\1/\2\comma\,\text{5}/\2}(-b\sqrt{t}) \right]
\end{align}
where the parameters $a$ and $b$ are offered by Eq. \eqref{eq:Eq10}. Equation \eqref{eq:Eq47} is a compact expression of the mean-square displacement in terms of the two-parameter Mittag--Leffler function and resembles the structure of its time-derivative given by Eq. \eqref{eq:Eq39new} and that of the velocity autocorrelation function given by Eq. \eqref{eq:Eq44new}. As we go up to higher-order time-derivatives of Eq. \eqref{eq:Eq47}, the power law of the time in front of the brackets reduces by one (from $t^{\nicefrac{\3}{\2}}$ to $t^{\nicefrac{\1}{\2}}$ to $t^{-\nicefrac{\1}{\2}}$) together with the second index of the Mittag--Leffler function (from $\nicefrac{\text{5}}{\2}$ to $\nicefrac{\3}{\2}$ to $\nicefrac{\1}{\2}$).

By employing the recurrence relation of the Mittag--Leffler function \citep{Erdelyi1953,GorenfloKilbasMainardiRogosin2014}:
\begin{equation}\label{eq:Eq39}
E_{p, q}(x)= \frac{\1}{x} E_{p, q-p}(x)-\frac{\1}{x \Gamma(q-p)}
\end{equation}
\begin{equation}\label{eq:Eq51}
E_{\nicefrac{\1}{\2}, \nicefrac{\text{5}}{\2}} \left( -at^{\nicefrac{\1}{\2}} \right) = - \frac{\1}{a\sqrt{t}} \left[ E_{\nicefrac{\1}{\2}, \2} \left( -a\sqrt{t} \right)-\1 \right]
\end{equation}
where $E_{\nicefrac{\1}{\2}, \2} \left( -a\sqrt{t} \right)$ in \eqref{eq:Eq51} is expressed as:
\begin{equation}\label{eq:Eq52}
E_{\nicefrac{\1}{\2}, \2} \left( -a\sqrt{t} \right) = - \frac{\1}{a\sqrt{t}} \left[ E_{\nicefrac{\1}{\2}, \nicefrac{\3}{\2}} \left( -a\sqrt{t} \right) - \frac{\1}{\Gamma \left(\nicefrac{\3}{\2}\right)} \right]
\end{equation}
Substitution of Eq. \eqref{eq:Eq52} into \eqref{eq:Eq51} after using that $\Gamma \left(\nicefrac{\3}{\2}\right) = \nicefrac{\sqrt{\pi}}{\2}$, an alternative expression of the mean-square displacement given by Eq. \eqref{eq:Eq47} is:
\begin{widetext}
\begin{equation}\label{eq:Eq53}
\left\langle \Delta r^\2(t) \right\rangle = \frac{N K_B T}{\3\pi R} J(t)= \frac{\2 N K_B T}{M} \left\lbrace \frac{t}{ab} +\frac{\sqrt{t}}{b-a}\frac{\2}{\sqrt{\pi}}\left( \frac{\1}{b^\2} -\frac{\1}{a^\2} \right) + \frac{\sqrt{t}}{b-a} \left[ \frac{\1}{a^{\2}} E_{\nicefrac{\1}{\2}, \nicefrac{\3}{\2}} \left( -a\sqrt{t} \right) - \frac{\1}{b^{\2}} E_{\nicefrac{\1}{\2}, \nicefrac{\3}{\2}} \left( -b\sqrt{t} \right) \right] \right\rbrace
\end{equation}
\end{widetext}
By employing once again the recurrence relation given by Eq. \eqref{eq:Eq39} in association with the identity:
\begin{equation}\label{eq:Eq40}
E_{\nicefrac{\1}{\2}, \1}(x)= e^{x^\2} \text{Erfc}(-x),
\end{equation}
the mean-square displacement of Brownian particles immersed in a Newtonian viscous fluid with hydrodynamic memory can be expressed as:
\begin{widetext}
\begin{equation}\label{eq:Eq54}
\left\langle \Delta r^\2(t) \right\rangle = \frac{N K_B T}{\3\pi R} J(t)= \frac{N K_B T}{\3\pi R} \frac{\1}{\eta} \left\lbrace t - \frac{\2}{\sqrt{\pi}} \frac{z}{\zeta} \sqrt{t} + \frac{z^\2 - \zeta M}{\zeta^\2} + \frac{\zeta}{M} \frac{\1}{b-a} \left[ \frac{\1}{b^{\3}} e^{b^\2 t} \text{Erfc}\left(b\sqrt{t}\right) - \frac{\1}{a^{\3}} e^{a^\2 t} \text{Erfc} \left(a\sqrt{t}\right) \right] \right\rbrace
\end{equation}
\end{widetext}
which is the classical expression known in the literature \citep{WeitzPinePuseyTough1989, ClercxSchram1992}. In deriving Eq. \eqref{eq:Eq54} we used the identities:
\begin{equation}\label{eq:Eq55}
\frac{a+b}{ab} = \frac{z}{\zeta} = \sqrt{\tau_f} \enskip \text{and} \enskip \frac{a^\2 + ab + b^\2}{a^\2 b^\2} = \frac{z^\2 - \zeta M}{\zeta^\2}
\end{equation}
which derive from the definitions of $a$ and $b$ given by Eq. \eqref{eq:Eq10}.

\begin{figure}[t!]
\centering
\includegraphics[width=\linewidth, angle=0]{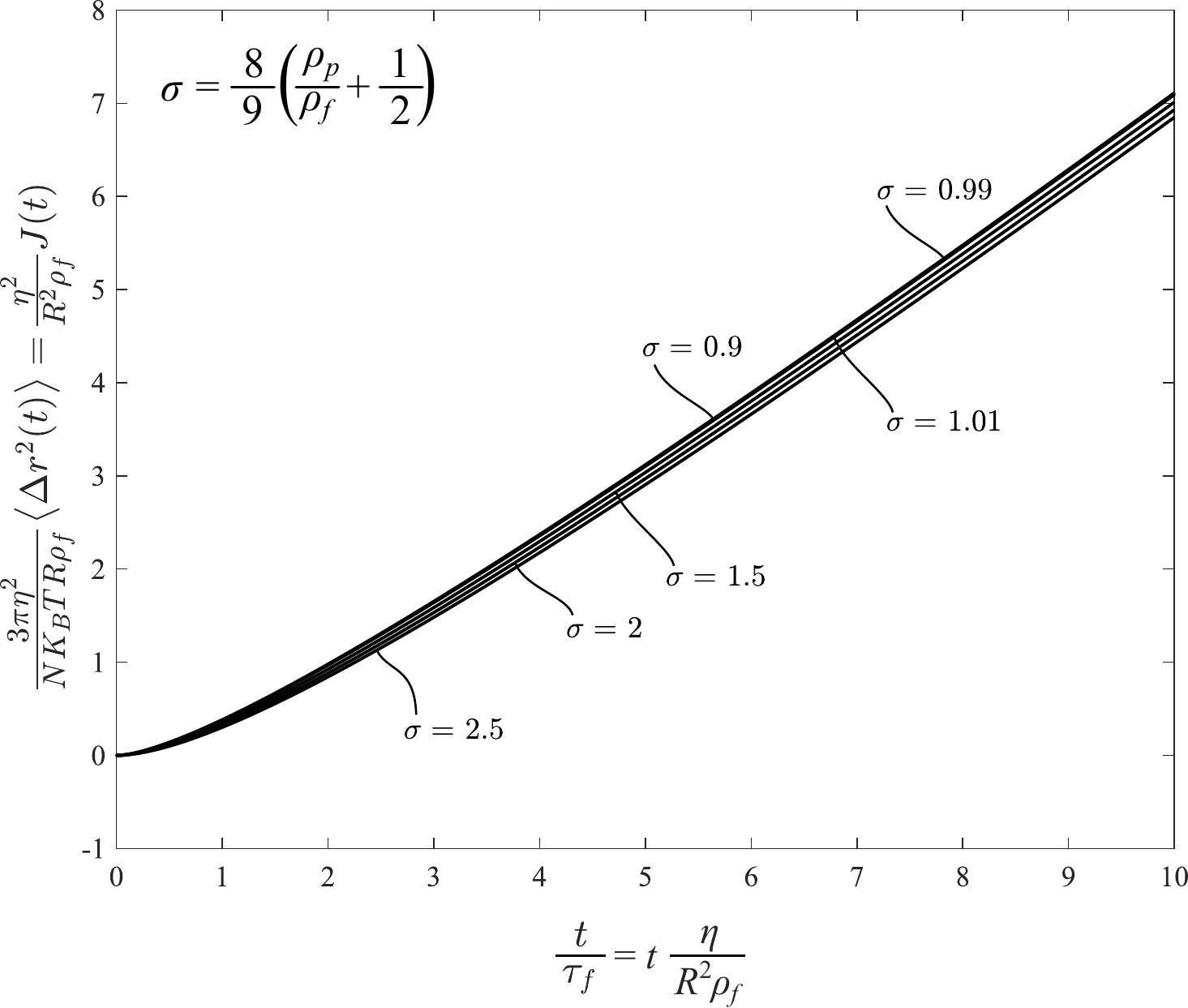}
\caption{Normalized mean-square displacement of Brownian particles immersed in a Newtonian, viscous fluid with viscosity $\eta$; when effects of hydrodynamic memory are accounted. The curves for $\sigma=\frac{\text{8}}{\text{9}}\left( \frac{\rho_p}{\rho_f} +\frac{\1}{\2}\right)=$ 0.9 and 0.99 are computed with Eq. \eqref{eq:Eq54new}; whereas, the curves for $\sigma=$ 1.01, 1.5, 2 and 2.5 are computed with Eq. \eqref{eq:Eq56} showing the continuity of the solution as parameter $\sigma$ crosses the value of $\sigma=\1$.}
\label{fig:Fig04}
\end{figure}

When $\frac{\rho_p}{\rho_f} > \frac{\text{5}}{\text{8}}$, parameter $\sigma > \1$ and the parameters $a$ and $b$ appearing in Eq. \eqref{eq:Eq54} are complex valued and are offered by Eq. \eqref{eq:Eq24}. In this case ($\sigma > \1$) Eq. \eqref{eq:Eq54} becomes:
\begin{widetext}
\begin{align}\label{eq:Eq56}
\left\langle \Delta r^\2(t) \right\rangle & = \frac{N K_B T}{\3\pi R} J(t)= \frac{N K_B T}{\3\pi R} \frac{\tau_f}{\eta} \left\lbrace \frac{t}{\tau_f} - \frac{\2}{\sqrt{\pi}} \sqrt{\frac{t}{\tau_f}} + \left( \1 - \frac{\sigma}{\4} \right) + \frac{\operatorname{i}}{\text{8}} \frac{\sigma^\3}{\sqrt{\sigma - \1}} e^{-\frac{\text{4}(\sigma -\2)}{\sigma^\2} \frac{t}{\tau_f}} \right. \\ \nonumber
& \times \left[ \frac{e^{-\frac{\operatorname{i}\text{8}\sqrt{\sigma -\1}}{\sigma^\2} \frac{t}{\tau_f}}}{\left( \1 - \operatorname{i}\sqrt{\sigma -\1} \right)^\3} \text{Erfc}\left( \frac{\2 \left( \1 - \operatorname{i}\sqrt{\sigma -\1} \right)}{\sigma} \sqrt{\frac{t}{\tau_f}} \right)
 \left. - \frac{e^{\frac{\operatorname{i}\text{8}\sqrt{\sigma -\1}}{\sigma^\2} \frac{t}{\tau_f}}}{\left( \1 + \operatorname{i}\sqrt{\sigma -\1} \right)^\3} \text{Erfc}\left( \frac{\2 \left( \1 + \operatorname{i}\sqrt{\sigma -\1} \right)}{\sigma} \sqrt{\frac{t}{\tau_f}} \right) \right] \right\rbrace, \enskip \sigma > \1
\end{align}
\end{widetext}
The mean-square displacement offered by Eq. \eqref{eq:Eq56} for $\sigma>$ 1 is a real-valued function given that the imaginary parts of the right-hand side of Eq. \eqref{eq:Eq56} cancel. For the case when $\sigma<\1$ $\left( \frac{\rho_p}{\rho_f} < \frac{\text{5}}{\text{8}} \right)$, the parameters $a$ and $b$ appearing in Eq. \eqref{eq:Eq54} are real-valued and are offered by Eq. \eqref{eq:Eq26}. In this case ($\sigma<$ 1), Eq. \eqref{eq:Eq54} becomes
\begin{widetext}
\begin{align}\label{eq:Eq54new}
\left\langle \Delta r^\2(t) \right\rangle & = \frac{N K_B T}{\3\pi R} J(t)= \frac{N K_B T}{\3\pi R} \frac{\tau_f}{\eta} \left\lbrace \frac{t}{\tau_f} - \frac{\2}{\sqrt{\pi}} \sqrt{\frac{t}{\tau_f}} + \left( \1 - \frac{\sigma}{\4} \right) - \frac{\sigma^\3}{\text{8}\sqrt{\1-\sigma}} e^{\frac{\text{4}(\sigma -\2)}{\sigma^\2} \frac{t}{\tau_f}} \right. \\ \nonumber
&  \times \left[ \frac{e^{-\frac{\text{8}\sqrt{\1-\sigma}}{\sigma^\2} \frac{t}{\tau_f}}}{\left( \1 - \sqrt{\1-\sigma } \right)^\3} \text{Erfc}\left( \frac{\2 \left( \1 - \sqrt{\1-\sigma } \right)}{\sigma} \sqrt{\frac{t}{\tau_f}} \right)
\left. - \frac{e^{\frac{\text{8}\sqrt{\sigma -\1}}{\sigma^\2} \frac{t}{\tau_f}}}{\left( \1 + \sqrt{\1-\sigma } \right)^\3} \text{Erfc}\left( \frac{\2 \left( \1 + \sqrt{\1-\sigma } \right)}{\sigma} \sqrt{\frac{t}{\tau_f}} \right) \right] \right\rbrace, \quad \sigma < \1
\end{align}
\end{widetext}
Figure \ref{fig:Fig04} plots the normalized mean-square displacement $\frac{\3\pi\eta^\2}{NK_BTR\rho_f}\left\langle \Delta r^\2(t) \right\rangle = \frac{\eta^\2}{R^\2 \rho_f}J(t)$ of Brownian particles with mass $m$ and radius $R$ immersed in a Newtonian, viscous fluid with viscosity $\eta$ for various values of the parameter $\sigma$ which accounts for hydrodynamic memory.

\section{Brownian Motion in Viscoelastic Materials with Hydrodynamic Memory}

The motion of a solid particle within a viscoelastic material is described by nonlinear continuum mechanics equations \citep{Gurtin1982,BirdArmstrongHassager1987} together with rheological models expressed in a corotating coordinate frame for constructing objective rheological equations of state. Nevertheless, when the particle motion happens over small distances in association with low Reynolds numbers, the nonlinearities that manifest in the dynamic equilibrium and constitutive equations can be neglected \citep{LandauLifshitz1959, BirdArmstrongHassager1987, ZwanzigBixon1970, ClercxSchram1992, LiverpoolMacKintosh2005, AtakhorramiMizunoKoenderinkLiverpoolMacKintoshSchmidt2008}. Accordingly, the unsteady motion of Brownian particles suspended in a linear, isotropic viscoelastic material is described by a generalized Langevin equation
\begin{equation}\label{eq:Eq55new}
m \frac{\mathrm{d}v(t)}{\mathrm{d}t} = -\int_{\0^-}^t \zeta (t-\xi)v(\xi) \, \mathrm{d}\xi + f_{R}(t)
\end{equation}
where $m$ is the particle mass, $v(t) = \frac{\mathrm{d}r(t)}{\mathrm{d}t}$ is the particle velocity and $f_{R}(t)$ are the random forces acting on the Brownian particle. The convolution integral in Eq. \eqref{eq:Eq55new} represents the drag force on the particle as it moves through the viscoelastic fluid over small distances and $\zeta (t-\xi)$ is the relaxation kernel of the Brownian particle--viscoelastic material system.

In their seminal paper Zwanzig and Bixon \cite{ZwanzigBixon1970} built upon Stoke's and Boussinesq's solution for the unsteady motion of a sphere within a Newtonian viscous fluid and upon decomposing the particle velocity into its Fourier components, $v(\omega) = \int_{-\infty}^\infty v(t) e^{-i\omega t} \, \mathrm{d}t$, reached a solution for the Fourier transform of the relaxation kernel $\mathcal{Z}(\omega) = \int_{-\infty}^\infty \zeta(t) e^{-i\omega t} \, \mathrm{d}t$. The mathematical expression for the impedance $\mathcal{Z}(\omega)$ derived by \cite{ZwanzigBixon1970} involves longitudinal and transverse Hankel functions after applying boundary conditions to the solution of a vector Helmholtz equation expressed in spherical harmonics. While the solution is valid for the general linear viscoelastic material, its complexity does not allow for closed form results other than at selective limited cases \citep{ZwanzigBixon1970}.

Clercx and Schram \cite{ClercxSchram1992} examined the motion of a Brownian particle suspended in a Newtonian viscous fluid under the influence of a harmonic potential that exerts a linear restoring force $kr(t)$, where $r(t)$ is the particle displacement and $k$ is the constant of proportionality (stiffness). With this configuration the drag force on the Brownian particle including the effects of backflow are associated solely with the surrounding viscous fluid appearing in the right-hand side of the generalized Langevin Eq. \eqref{eq:Eq04}, or of its alternative form given by Eq. \eqref{eq:Eq08}.

With reference to Eq. \eqref{eq:Eq08} and after expressing the particle velocity $v(t) = \frac{\mathrm{d}r(t)}{\mathrm{d}t}$, the generalized Langevin equation for the Brownian particle in a harmonic trap (Kelvin-Voigt solid) when effects from hydrodynamic memory are accounted simplifies to
\begin{equation}\label{eq:Eq56new}
M \frac{\mathrm{d}^\2 r(t)}{\mathrm{d}t^\2} + \text{6} \pi R^\2 \sqrt{\rho_{f} \eta} \frac{\mathrm{d}^{\nicefrac{\3}{\2}} r(t)}{\mathrm{d}t^{\nicefrac{\3}{\2}}} + \text{6} \pi R \eta \frac{\mathrm{d} r(t)}{\mathrm{d}t} + k r(t) = f_{R} (t)
\end{equation}
where again $M = \frac{\4}{\3} \pi R^\3 (\rho_{p} + \frac{\1}{\2}\rho_{f}) = \frac{\3}{\2} \pi R^\3 \rho_{f} \sigma$ with $\sigma$ offered by Eq. \eqref{eq:Eq22}. Clercx and Schram \cite{ClercxSchram1992}, after deriving the appropriate correlation function $\left\langle f_{R}(t_{\1}) f_{R}(t_{\2})\right\rangle$ of the random forces, derived closed form expressions of the velocity autocorrelation function and of the mean-square displacements by computing ensemble averages of the random Brownian process described by Eq. \eqref{eq:Eq56new}.

\begin{figure}[t!]
\centering
\includegraphics[width=0.9\linewidth, angle=0]{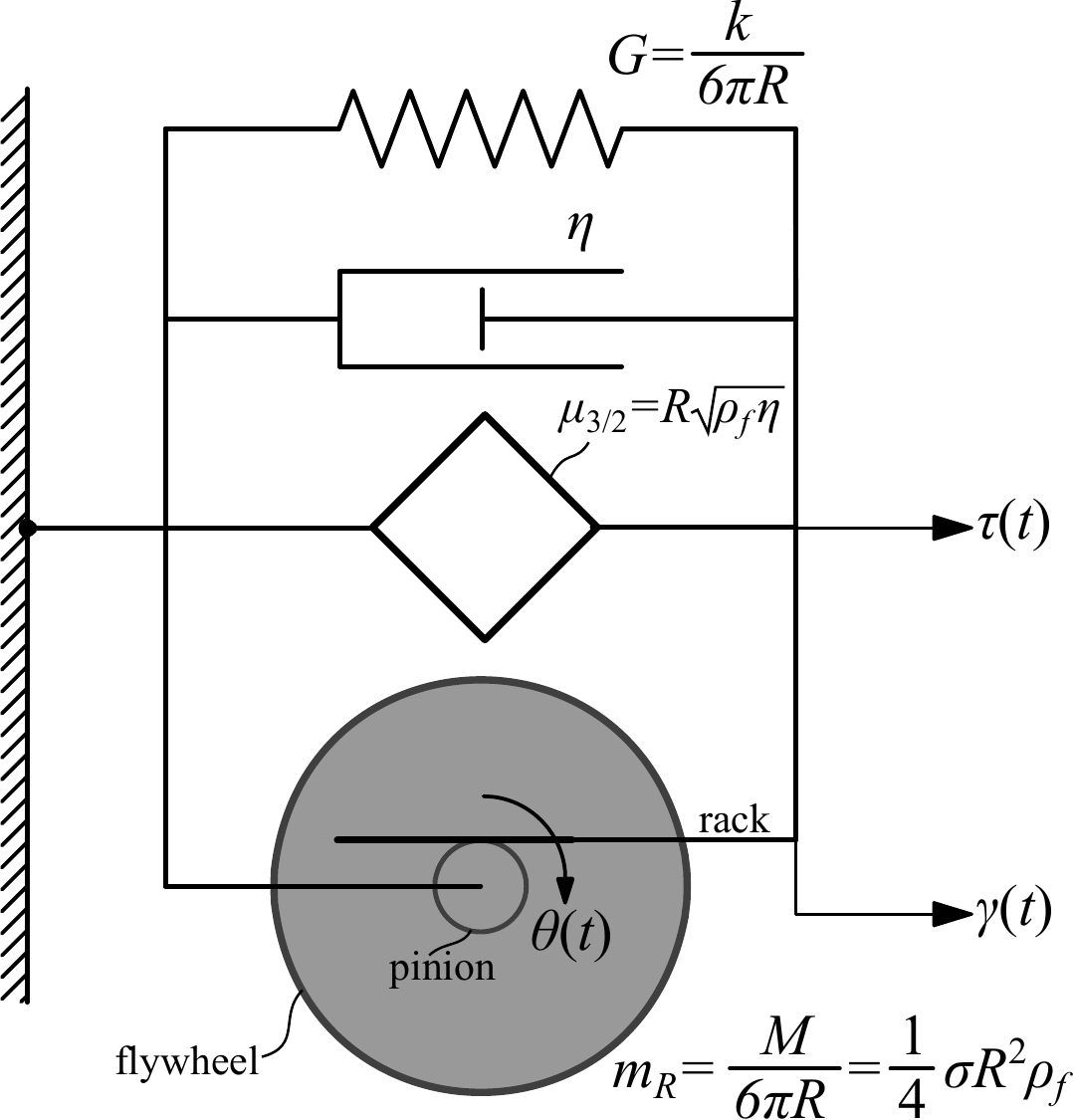}
\caption{The rheological analogue for Brownian motion in a viscously damped harmonic trap with hydrodynamic memory. A spring with shear modulus $G$, a dashpot with viscosity $\eta$, a Scott-Blair element with material constant $\mu_{\nicefrac{\3}{\2}}=R\sqrt{\rho_f \eta}$ and an inerter with distributed inertance $m_R=\frac{M}{\6\pi R}=\frac{\2}{\text{9}}R^\2\rho_f\left( \frac{\rho_p}{\rho_f} + \frac{\1}{\2}\right)=\frac{\1}{\text{4}}\sigma R^\2 \rho_f$ are connected in parallel; the spring--dashpot--inertpot--inerter parallel connection.}
\label{fig:Fig05}
\end{figure}

The concept of a rheological analogue advanced in this paper can be also applied for the case of Brownian motion in a harmonic trap (Kelvin-Voigt solid) when hydrodynamic memory is accounted. The synthesis of the proposed macroscopic mechanical network is suggested from the left-hand side of the Langevin equation \eqref{eq:Eq56new}, which consists of a inertia term; a $\nicefrac{\3}{\2}$-order fractional derivative of the displacement term, a viscous term and an elastic term acting in parallel to balance the random force $f_{R}(t)$. Accordingly, by applying the viscous--viscoelastic correspondence principle for Brownian motion \citep{Makris2020, Makris2021SM} the mean-square displacement and its time derivatives of Brownian particles in a harmonic trap when hydrodynamic memory is accounted are evaluated by merely computing the creep compliance, $J(t) = \frac{\3 \pi R}{N K_{B} T} \left\langle \Delta r^\2(t) \right\rangle$, impulse response function, $h(t) = \frac{\3 \pi R}{N K_{B} T} \frac{\mathrm{d}\left\langle \Delta r^\2(t) \right\rangle}{\mathrm{d}t}$ and impulse strain rate response function $\psi(t) = \frac{\3 \pi R}{N K_{B} T} \frac{\mathrm{d}^\2\left\langle \Delta r^\2(t) \right\rangle}{\mathrm{d}t^\2} = \frac{\text{6} \pi R}{N K_{B} T} \left\langle v(\0) v(t) \right\rangle$ of the mechanical network shown in Fig. \ref{fig:Fig05}. Given the parallel connection of the spring--dashpot--interpot--inerter network shown in Fig. \ref{fig:Fig05}, its constitutive law is
\begin{equation}\label{eq:Eq57new}
\tau(t) = G \gamma(t) + \eta \frac{\mathrm{d}\gamma(t)}{\mathrm{d}t} + \mu_{\nicefrac{\3}{\2}} \frac{\mathrm{d}^{\nicefrac{\3}{\2}}\gamma(t)}{\mathrm{d}t^{\nicefrac{\3}{\2}}} + m_{R} \frac{\mathrm{d}^\2\gamma(t)}{\mathrm{d}t^\2}
\end{equation}
The Laplace transform of Eq. \eqref{eq:Eq57new} gives
\begin{equation}\label{eq:Eq58new}
\tau(s) = \mathcal{G}(s) \gamma(s) =(G + \eta s + \mu_{\nicefrac{\3}{\2}} s^{\nicefrac{\3}{\2}} + m_{R} s^\2) \gamma(s)
\end{equation}
where $\mathcal{G}(s) = G + \eta s + \mu_{\nicefrac{\3}{\2}} s^{\nicefrac{\3}{\2}} + m_{R} s^\2$ is the complex dynamic modulus. The complex dynamic compliance of the rheological network shown in Fig. \ref{fig:Fig05} is
\begin{align}\label{eq:Eq59new}
\mathcal{J}(s) & = \frac{\1}{\mathcal{G}(s)} = \frac{\1}{(G + \eta s + \mu_{\nicefrac{\3}{\2}} s^{\nicefrac{\3}{\2}} + m_{R} s^\2)} \\ \nonumber 
& = \frac{\1}{m_{R}} \frac{\1}{s^\2 + \frac{\mu_{\nicefrac{\3}{\2}}}{m_{R}} s^{\nicefrac{\3}{\2}} + \frac{\eta}{m_{R}} s + \omega^\2_{R}}
\end{align}
where $\omega_{R} = \sqrt{\frac{G}{m_{R}}} = \sqrt{\frac{k}{M}}$ is the undamped natural frequency of the trapped Brownian particle. The Laplace transform of the velocity autocorrelation function of Brownian particles in a harmonic trap when hydrodynamic memory is accounted is offered by Eq. \eqref{eq:Eq15} where now the complex dynamic compliance $\mathcal{J}(s)$ is offered by Eq. \eqref{eq:Eq59new}. Inverse Laplace transform of Eq. \eqref{eq:Eq15} yields the evolved expression for the velocity autocorrelation function first presented by \cite{ClercxSchram1992}.

\section{Summary}

In this paper we first recognized that for Brownian particles immersed in a Newtonian viscous fluid, the hydrodynamic memory term associated with the backflow is essentially a $\nicefrac{\1}{\2}$-fractional derivative of the velocity of the Brownian particle. Accordingly, the hydrodynamic memory term can be expressed with a macroscopic Scott-Blair fractional element. Subsequently, we built on a recently published viscous--viscoelastic correspondence principle for Brownian motion \citep{Makris2020, Makris2021SM} and presented macroscopic rheological networks for Brownian motion of particles either immersed in a Newtonian viscous fluid or trapped in a harmonic potential. The synthesis of the rheological network is suggested from the structure of the generalized Langevin equation and simplifies appreciably the calculations for evaluating the mean-square displacement and its time derivatives.

\bibliographystyle{apsrev4-2.bst}
\bibliography{References}

\newcommand{\noopsort}[1]{}
\begin{thebibliography}{57}%
\makeatletter
\providecommand \@ifxundefined [1]{%
 \@ifx{#1\undefined}
}%
\providecommand \@ifnum [1]{%
 \ifnum #1\expandafter \@firstoftwo
 \else \expandafter \@secondoftwo
 \fi
}%
\providecommand \@ifx [1]{%
 \ifx #1\expandafter \@firstoftwo
 \else \expandafter \@secondoftwo
 \fi
}%
\providecommand \natexlab [1]{#1}%
\providecommand \enquote  [1]{``#1''}%
\providecommand \bibnamefont  [1]{#1}%
\providecommand \bibfnamefont [1]{#1}%
\providecommand \citenamefont [1]{#1}%
\providecommand \href@noop [0]{\@secondoftwo}%
\providecommand \href [0]{\begingroup \@sanitize@url \@href}%
\providecommand \@href[1]{\@@startlink{#1}\@@href}%
\providecommand \@@href[1]{\endgroup#1\@@endlink}%
\providecommand \@sanitize@url [0]{\catcode `\\12\catcode `\$12\catcode
  `\&12\catcode `\#12\catcode `\^12\catcode `\_12\catcode `\%12\relax}%
\providecommand \@@startlink[1]{}%
\providecommand \@@endlink[0]{}%
\providecommand \url  [0]{\begingroup\@sanitize@url \@url }%
\providecommand \@url [1]{\endgroup\@href {#1}{\urlprefix }}%
\providecommand \urlprefix  [0]{URL }%
\providecommand \Eprint [0]{\href }%
\providecommand \doibase [0]{https://doi.org/}%
\providecommand \selectlanguage [0]{\@gobble}%
\providecommand \bibinfo  [0]{\@secondoftwo}%
\providecommand \bibfield  [0]{\@secondoftwo}%
\providecommand \translation [1]{[#1]}%
\providecommand \BibitemOpen [0]{}%
\providecommand \bibitemStop [0]{}%
\providecommand \bibitemNoStop [0]{.\EOS\space}%
\providecommand \EOS [0]{\spacefactor3000\relax}%
\providecommand \BibitemShut  [1]{\csname bibitem#1\endcsname}%
\let\auto@bib@innerbib\@empty
\bibitem [{\citenamefont {Einstein}(1905)}]{Einstein1905}%
  \BibitemOpen
  \bibfield  {author} {\bibinfo {author} {\bibfnamefont {A.}~\bibnamefont
  {Einstein}},\ }\href@noop {} {\bibfield  {journal} {\bibinfo  {journal}
  {Annalen der Physik}\ }\textbf {\bibinfo {volume} {17}},\ \bibinfo {pages}
  {549} (\bibinfo {year} {1905})}\BibitemShut {NoStop}%
\bibitem [{\citenamefont {Langevin}(1908)}]{Langevin1908}%
  \BibitemOpen
  \bibfield  {author} {\bibinfo {author} {\bibfnamefont {P.}~\bibnamefont
  {Langevin}},\ }\href@noop {} {\bibfield  {journal} {\bibinfo  {journal}
  {Compt. Rendus}\ }\textbf {\bibinfo {volume} {146}},\ \bibinfo {pages} {530}
  (\bibinfo {year} {1908})}\BibitemShut {NoStop}%
\bibitem [{\citenamefont {Landau}\ and\ \citenamefont
  {Lifshitz}(1959)}]{LandauLifshitz1959}%
  \BibitemOpen
  \bibfield  {author} {\bibinfo {author} {\bibfnamefont {L.~D.}\ \bibnamefont
  {Landau}}\ and\ \bibinfo {author} {\bibfnamefont {E.~M.}\ \bibnamefont
  {Lifshitz}},\ }\href@noop {} {\emph {\bibinfo {title} {Course of Theoretical
  Physics Vol. 6 Fluid Mechanies}}}\ (\bibinfo  {publisher} {Pergamon Press},\
  \bibinfo {year} {1959})\BibitemShut {NoStop}%
\bibitem [{\citenamefont {Attard}(2012)}]{Attard2012}%
  \BibitemOpen
  \bibfield  {author} {\bibinfo {author} {\bibfnamefont {P.}~\bibnamefont
  {Attard}},\ }\href@noop {} {\emph {\bibinfo {title} {Non-equilibrium
  thermodynamics and statistical mechanics: Foundations and applications}}}\
  (\bibinfo  {publisher} {OUP Oxford},\ \bibinfo {year} {2012})\BibitemShut
  {NoStop}%
\bibitem [{\citenamefont {Kalmykov}\ and\ \citenamefont
  {Coffey}(2017)}]{KalmykovCoffey2017}%
  \BibitemOpen
  \bibfield  {author} {\bibinfo {author} {\bibfnamefont {Y.~P.}\ \bibnamefont
  {Kalmykov}}\ and\ \bibinfo {author} {\bibfnamefont {W.~T.}\ \bibnamefont
  {Coffey}},\ }\href@noop {} {\emph {\bibinfo {title} {The Langevin
  Equation}}}\ (\bibinfo  {publisher} {World Scientific Publishing Company},\
  \bibinfo {year} {2017})\BibitemShut {NoStop}%
\bibitem [{\citenamefont {Uhlenbeck}\ and\ \citenamefont
  {Ornstein}(1930)}]{UhlenbeckOrnstein1930}%
  \BibitemOpen
  \bibfield  {author} {\bibinfo {author} {\bibfnamefont {G.~E.}\ \bibnamefont
  {Uhlenbeck}}\ and\ \bibinfo {author} {\bibfnamefont {L.~S.}\ \bibnamefont
  {Ornstein}},\ }\href@noop {} {\bibfield  {journal} {\bibinfo  {journal}
  {Physical Review}\ }\textbf {\bibinfo {volume} {36}},\ \bibinfo {pages} {823}
  (\bibinfo {year} {1930})}\BibitemShut {NoStop}%
\bibitem [{\citenamefont {Wang}\ and\ \citenamefont
  {Uhlenbeck}(1945)}]{WangUhlenbeck1945}%
  \BibitemOpen
  \bibfield  {author} {\bibinfo {author} {\bibfnamefont {M.~C.}\ \bibnamefont
  {Wang}}\ and\ \bibinfo {author} {\bibfnamefont {G.~E.}\ \bibnamefont
  {Uhlenbeck}},\ }\href@noop {} {\bibfield  {journal} {\bibinfo  {journal}
  {Reviews of Modern Physics}\ }\textbf {\bibinfo {volume} {17}},\ \bibinfo
  {pages} {323} (\bibinfo {year} {1945})}\BibitemShut {NoStop}%
\bibitem [{\citenamefont {Lighthill}(1958)}]{Lighthill1958}%
  \BibitemOpen
  \bibfield  {author} {\bibinfo {author} {\bibfnamefont {M.~J.}\ \bibnamefont
  {Lighthill}},\ }\href@noop {} {\emph {\bibinfo {title} {An introduction to
  Fourier analysis and generalised functions}}}\ (\bibinfo  {publisher}
  {Cambridge University Press},\ \bibinfo {year} {1958})\BibitemShut {NoStop}%
\bibitem [{\citenamefont {Zwanzig}\ and\ \citenamefont
  {Bixon}(1970)}]{ZwanzigBixon1970}%
  \BibitemOpen
  \bibfield  {author} {\bibinfo {author} {\bibfnamefont {R.}~\bibnamefont
  {Zwanzig}}\ and\ \bibinfo {author} {\bibfnamefont {M.}~\bibnamefont
  {Bixon}},\ }\href@noop {} {\bibfield  {journal} {\bibinfo  {journal}
  {Physical Review A}\ }\textbf {\bibinfo {volume} {2}},\ \bibinfo {pages}
  {2005} (\bibinfo {year} {1970})}\BibitemShut {NoStop}%
\bibitem [{\citenamefont {Widom}(1971)}]{Widom1971}%
  \BibitemOpen
  \bibfield  {author} {\bibinfo {author} {\bibfnamefont {A.}~\bibnamefont
  {Widom}},\ }\href@noop {} {\bibfield  {journal} {\bibinfo  {journal}
  {Physical Review A}\ }\textbf {\bibinfo {volume} {3}},\ \bibinfo {pages}
  {1394} (\bibinfo {year} {1971})}\BibitemShut {NoStop}%
\bibitem [{\citenamefont {Hinch}(1975)}]{Hinch1975}%
  \BibitemOpen
  \bibfield  {author} {\bibinfo {author} {\bibfnamefont {E.~J.}\ \bibnamefont
  {Hinch}},\ }\href@noop {} {\bibfield  {journal} {\bibinfo  {journal} {Journal
  of Fluid Mechanics}\ }\textbf {\bibinfo {volume} {72}},\ \bibinfo {pages}
  {499} (\bibinfo {year} {1975})}\BibitemShut {NoStop}%
\bibitem [{\citenamefont {Paul}\ and\ \citenamefont
  {Pusey}(1981)}]{PaulPusey1981}%
  \BibitemOpen
  \bibfield  {author} {\bibinfo {author} {\bibfnamefont {G.~L.}\ \bibnamefont
  {Paul}}\ and\ \bibinfo {author} {\bibfnamefont {P.~N.}\ \bibnamefont
  {Pusey}},\ }\href@noop {} {\bibfield  {journal} {\bibinfo  {journal} {Journal
  of Physics A: Mathematical and General}\ }\textbf {\bibinfo {volume} {14}},\
  \bibinfo {pages} {3301} (\bibinfo {year} {1981})}\BibitemShut {NoStop}%
\bibitem [{\citenamefont {Clercx}\ and\ \citenamefont
  {Schram}(1992)}]{ClercxSchram1992}%
  \BibitemOpen
  \bibfield  {author} {\bibinfo {author} {\bibfnamefont {H.~J.~H.}\
  \bibnamefont {Clercx}}\ and\ \bibinfo {author} {\bibfnamefont {P.~P. J.~M.}\
  \bibnamefont {Schram}},\ }\href@noop {} {\bibfield  {journal} {\bibinfo
  {journal} {Physical Review A}\ }\textbf {\bibinfo {volume} {46}},\ \bibinfo
  {pages} {1942} (\bibinfo {year} {1992})}\BibitemShut {NoStop}%
\bibitem [{\citenamefont {Oldham}\ and\ \citenamefont
  {Spanier}(1974)}]{OldhamSpanier1974}%
  \BibitemOpen
  \bibfield  {author} {\bibinfo {author} {\bibfnamefont {K.}~\bibnamefont
  {Oldham}}\ and\ \bibinfo {author} {\bibfnamefont {J.}~\bibnamefont
  {Spanier}},\ }\href@noop {} {\emph {\bibinfo {title} {The Fractional
  Calculus. Mathematics in science and engineering}}},\ Vol.\ \bibinfo {volume}
  {III}\ (\bibinfo  {publisher} {Academic Press Inc.},\ \bibinfo {address} {San
  Diego, CA},\ \bibinfo {year} {1974})\BibitemShut {NoStop}%
\bibitem [{\citenamefont {Samko}\ \emph {et~al.}(1974)\citenamefont {Samko},
  \citenamefont {Kilbas},\ and\ \citenamefont
  {Marichev}}]{SamkoKilbasMarichev1974}%
  \BibitemOpen
  \bibfield  {author} {\bibinfo {author} {\bibfnamefont {S.~G.}\ \bibnamefont
  {Samko}}, \bibinfo {author} {\bibfnamefont {A.~A.}\ \bibnamefont {Kilbas}},\
  and\ \bibinfo {author} {\bibfnamefont {O.~I.}\ \bibnamefont {Marichev}},\
  }\href@noop {} {\emph {\bibinfo {title} {Fractional Integrals and
  Derivatives; Theory and Applications}}},\ Vol.~\bibinfo {volume} {1}\
  (\bibinfo  {publisher} {Gordon and Breach Science Publishers},\ \bibinfo
  {address} {Amsterdam},\ \bibinfo {year} {1974})\BibitemShut {NoStop}%
\bibitem [{\citenamefont {Miller}\ and\ \citenamefont
  {Ross}(1993)}]{MillerRoss1993}%
  \BibitemOpen
  \bibfield  {author} {\bibinfo {author} {\bibfnamefont {K.~S.}\ \bibnamefont
  {Miller}}\ and\ \bibinfo {author} {\bibfnamefont {B.}~\bibnamefont {Ross}},\
  }\href@noop {} {\emph {\bibinfo {title} {An introduction to the fractional
  calculus and fractional differential equations}}}\ (\bibinfo  {publisher}
  {Wiley},\ \bibinfo {address} {New York, NY},\ \bibinfo {year}
  {1993})\BibitemShut {NoStop}%
\bibitem [{\citenamefont {Podlubny}(1998)}]{Podlubny1998}%
  \BibitemOpen
  \bibfield  {author} {\bibinfo {author} {\bibfnamefont {I.}~\bibnamefont
  {Podlubny}},\ }\href@noop {} {\emph {\bibinfo {title} {Fractional
  differential equations: An introduction to fractional derivatives, fractional
  differential equations, to methods of their solution and some of their
  applications}}}\ (\bibinfo  {publisher} {Elsevier},\ \bibinfo {year}
  {1998})\BibitemShut {NoStop}%
\bibitem [{\citenamefont {Mainardi}(2010)}]{Mainardi2010}%
  \BibitemOpen
  \bibfield  {author} {\bibinfo {author} {\bibfnamefont {F.}~\bibnamefont
  {Mainardi}},\ }\href@noop {} {\emph {\bibinfo {title} {Fractional calculus
  and waves in linear viscoelasticity: An introduction to mathematical
  models}}}\ (\bibinfo  {publisher} {Imperial College Press - World
  Scientific},\ \bibinfo {address} {London, UK},\ \bibinfo {year}
  {2010})\BibitemShut {NoStop}%
\bibitem [{\citenamefont {Makris}(2021{\natexlab{a}})}]{Makris2021a}%
  \BibitemOpen
  \bibfield  {author} {\bibinfo {author} {\bibfnamefont {N.}~\bibnamefont
  {Makris}},\ }\href@noop {} {\bibfield  {journal} {\bibinfo  {journal}
  {Fractal and Fractional}\ }\textbf {\bibinfo {volume} {5}},\ \bibinfo {pages}
  {18} (\bibinfo {year} {2021}{\natexlab{a}})}\BibitemShut {NoStop}%
\bibitem [{\citenamefont {Nutting}(1921)}]{Nutting1921}%
  \BibitemOpen
  \bibfield  {author} {\bibinfo {author} {\bibfnamefont {P.~G.}\ \bibnamefont
  {Nutting}},\ }\href@noop {} {\bibfield  {journal} {\bibinfo  {journal}
  {Proceedings American Society for Testing Materials}\ }\textbf {\bibinfo
  {volume} {21}},\ \bibinfo {pages} {1162} (\bibinfo {year}
  {1921})}\BibitemShut {NoStop}%
\bibitem [{\citenamefont {Gemant}(1936)}]{Gemant1936}%
  \BibitemOpen
  \bibfield  {author} {\bibinfo {author} {\bibfnamefont {A.}~\bibnamefont
  {Gemant}},\ }\href@noop {} {\bibfield  {journal} {\bibinfo  {journal}
  {Physics}\ }\textbf {\bibinfo {volume} {7}},\ \bibinfo {pages} {311}
  (\bibinfo {year} {1936})}\BibitemShut {NoStop}%
\bibitem [{\citenamefont {Scott~Blair}(1944)}]{ScottBlair1944}%
  \BibitemOpen
  \bibfield  {author} {\bibinfo {author} {\bibfnamefont {G.~W.}\ \bibnamefont
  {Scott~Blair}},\ }\href@noop {} {\emph {\bibinfo {title} {A survey of general
  and applied rheology}}}\ (\bibinfo  {publisher} {Isaac Pitman \& Sons},\
  \bibinfo {year} {1944})\BibitemShut {NoStop}%
\bibitem [{\citenamefont {Scott~Blair}(1947)}]{ScottBlair1947}%
  \BibitemOpen
  \bibfield  {author} {\bibinfo {author} {\bibfnamefont {G.~W.}\ \bibnamefont
  {Scott~Blair}},\ }\href@noop {} {\bibfield  {journal} {\bibinfo  {journal}
  {Journal of Colloid Science}\ }\textbf {\bibinfo {volume} {2}},\ \bibinfo
  {pages} {21} (\bibinfo {year} {1947})}\BibitemShut {NoStop}%
\bibitem [{\citenamefont {Li}\ and\ \citenamefont
  {Raizen}(2013)}]{LiRaizen2013}%
  \BibitemOpen
  \bibfield  {author} {\bibinfo {author} {\bibfnamefont {T.}~\bibnamefont
  {Li}}\ and\ \bibinfo {author} {\bibfnamefont {M.~G.}\ \bibnamefont
  {Raizen}},\ }\href@noop {} {\bibfield  {journal} {\bibinfo  {journal}
  {Annalen der Physik}\ }\textbf {\bibinfo {volume} {525}},\ \bibinfo {pages}
  {281} (\bibinfo {year} {2013})}\BibitemShut {NoStop}%
\bibitem [{\citenamefont {Makris}(2020)}]{Makris2020}%
  \BibitemOpen
  \bibfield  {author} {\bibinfo {author} {\bibfnamefont {N.}~\bibnamefont
  {Makris}},\ }\href@noop {} {\bibfield  {journal} {\bibinfo  {journal}
  {Physical Review E}\ }\textbf {\bibinfo {volume} {101}},\ \bibinfo {pages}
  {052139} (\bibinfo {year} {2020})}\BibitemShut {NoStop}%
\bibitem [{\citenamefont {Makris}(2021{\natexlab{b}})}]{Makris2021SM}%
  \BibitemOpen
  \bibfield  {author} {\bibinfo {author} {\bibfnamefont {N.}~\bibnamefont
  {Makris}},\ }\href@noop {} {\bibfield  {journal} {\bibinfo  {journal} {Soft
  Matter}\ }\textbf {\bibinfo {volume} {17}},\ \bibinfo {pages} {5410}
  (\bibinfo {year} {2021}{\natexlab{b}})}\BibitemShut {NoStop}%
\bibitem [{\citenamefont {Gemant}(1938)}]{Gemant1938}%
  \BibitemOpen
  \bibfield  {author} {\bibinfo {author} {\bibfnamefont {A.}~\bibnamefont
  {Gemant}},\ }\href@noop {} {\bibfield  {journal} {\bibinfo  {journal} {The
  London, Edinburgh, and Dublin Philosophical Magazine and Journal of Science}\
  }\textbf {\bibinfo {volume} {25}},\ \bibinfo {pages} {540} (\bibinfo {year}
  {1938})}\BibitemShut {NoStop}%
\bibitem [{\citenamefont {Smith}(2002)}]{Smith2002}%
  \BibitemOpen
  \bibfield  {author} {\bibinfo {author} {\bibfnamefont {M.~C.}\ \bibnamefont
  {Smith}},\ }\href@noop {} {\bibfield  {journal} {\bibinfo  {journal} {IEEE
  Transactions on Automatic Control}\ }\textbf {\bibinfo {volume} {47}},\
  \bibinfo {pages} {1648} (\bibinfo {year} {2002})}\BibitemShut {NoStop}%
\bibitem [{\citenamefont {Papageorgiou}\ and\ \citenamefont
  {Smith}(2005)}]{PapageorgiouSmith2005}%
  \BibitemOpen
  \bibfield  {author} {\bibinfo {author} {\bibfnamefont {C.}~\bibnamefont
  {Papageorgiou}}\ and\ \bibinfo {author} {\bibfnamefont {M.~C.}\ \bibnamefont
  {Smith}},\ }in\ \href@noop {} {\emph {\bibinfo {booktitle} {Proceedings of
  the 44th IEEE Conference on Decision and Control}}}\ (\bibinfo {organization}
  {IEEE},\ \bibinfo {year} {2005})\ pp.\ \bibinfo {pages}
  {3351--3356}\BibitemShut {NoStop}%
\bibitem [{\citenamefont {Makris}(2017)}]{Makris2017}%
  \BibitemOpen
  \bibfield  {author} {\bibinfo {author} {\bibfnamefont {N.}~\bibnamefont
  {Makris}},\ }\href@noop {} {\bibfield  {journal} {\bibinfo  {journal}
  {Journal of Engineering Mechanics}\ }\textbf {\bibinfo {volume} {143}},\
  \bibinfo {pages} {04017123} (\bibinfo {year} {2017})}\BibitemShut {NoStop}%
\bibitem [{\citenamefont {Makris}(2018)}]{Makris2018}%
  \BibitemOpen
  \bibfield  {author} {\bibinfo {author} {\bibfnamefont {N.}~\bibnamefont
  {Makris}},\ }\href@noop {} {\bibfield  {journal} {\bibinfo  {journal}
  {Meccanica}\ }\textbf {\bibinfo {volume} {53}},\ \bibinfo {pages} {2237}
  (\bibinfo {year} {2018})}\BibitemShut {NoStop}%
\bibitem [{\citenamefont {Erd{\'e}lyi}(1954)}]{Erdelyi1954}%
  \BibitemOpen
  \bibinfo {editor} {\bibfnamefont {A.}~\bibnamefont {Erd{\'e}lyi}},\ ed.,\
  \href@noop {} {\emph {\bibinfo {title} {Bateman Manuscript Project, Tables of
  Integral Transforms Vol I}}}\ (\bibinfo  {publisher} {McGraw-Hill},\ \bibinfo
  {address} {New York, NY},\ \bibinfo {year} {1954})\BibitemShut {NoStop}%
\bibitem [{\citenamefont {Erd{\'e}lyi}(1953)}]{Erdelyi1953}%
  \BibitemOpen
  \bibinfo {editor} {\bibfnamefont {A.}~\bibnamefont {Erd{\'e}lyi}},\ ed.,\
  \href@noop {} {\emph {\bibinfo {title} {Bateman Manuscript Project, Higher
  Transcendental Functions Vol III}}}\ (\bibinfo  {publisher} {McGraw-Hill},\
  \bibinfo {address} {New York, NY},\ \bibinfo {year} {1953})\BibitemShut
  {NoStop}%
\bibitem [{\citenamefont {Gorenflo}\ \emph {et~al.}(2014)\citenamefont
  {Gorenflo}, \citenamefont {Kilbas}, \citenamefont {Mainardi}, \citenamefont
  {Rogosin} \emph {et~al.}}]{GorenfloKilbasMainardiRogosin2014}%
  \BibitemOpen
  \bibfield  {author} {\bibinfo {author} {\bibfnamefont {R.}~\bibnamefont
  {Gorenflo}}, \bibinfo {author} {\bibfnamefont {A.~A.}\ \bibnamefont
  {Kilbas}}, \bibinfo {author} {\bibfnamefont {F.}~\bibnamefont {Mainardi}},
  \bibinfo {author} {\bibfnamefont {S.~V.}\ \bibnamefont {Rogosin}}, \emph
  {et~al.},\ }\href@noop {} {\emph {\bibinfo {title} {Mittag-Leffler functions,
  related topics and applications Vol. 2}}}\ (\bibinfo  {publisher}
  {Springer},\ \bibinfo {year} {2014})\BibitemShut {NoStop}%
\bibitem [{\citenamefont {Franosch}\ \emph {et~al.}(2011)\citenamefont
  {Franosch}, \citenamefont {Grimm}, \citenamefont {Belushkin}, \citenamefont
  {Mor}, \citenamefont {Foffi}, \citenamefont {Forr{\'o}},\ and\ \citenamefont
  {Jeney}}]{Franosch_etal2011}%
  \BibitemOpen
  \bibfield  {author} {\bibinfo {author} {\bibfnamefont {T.}~\bibnamefont
  {Franosch}}, \bibinfo {author} {\bibfnamefont {M.}~\bibnamefont {Grimm}},
  \bibinfo {author} {\bibfnamefont {M.}~\bibnamefont {Belushkin}}, \bibinfo
  {author} {\bibfnamefont {F.~M.}\ \bibnamefont {Mor}}, \bibinfo {author}
  {\bibfnamefont {G.}~\bibnamefont {Foffi}}, \bibinfo {author} {\bibfnamefont
  {L.}~\bibnamefont {Forr{\'o}}},\ and\ \bibinfo {author} {\bibfnamefont
  {S.}~\bibnamefont {Jeney}},\ }\href@noop {} {\bibfield  {journal} {\bibinfo
  {journal} {Nature}\ }\textbf {\bibinfo {volume} {478}},\ \bibinfo {pages}
  {85} (\bibinfo {year} {2011})}\BibitemShut {NoStop}%
\bibitem [{\citenamefont {Dom{\'i}nguez-Garc{\'i}a}\ \emph
  {et~al.}(2014)\citenamefont {Dom{\'i}nguez-Garc{\'i}a}, \citenamefont
  {Cardinaux}, \citenamefont {Bertseva}, \citenamefont {Forr{\'o}},
  \citenamefont {Scheffold},\ and\ \citenamefont
  {Jeney}}]{DominguezGarciaCardinauxBertsevaForroScheffoldJeney2014}%
  \BibitemOpen
  \bibfield  {author} {\bibinfo {author} {\bibfnamefont {P.}~\bibnamefont
  {Dom{\'i}nguez-Garc{\'i}a}}, \bibinfo {author} {\bibfnamefont
  {F.}~\bibnamefont {Cardinaux}}, \bibinfo {author} {\bibfnamefont
  {E.}~\bibnamefont {Bertseva}}, \bibinfo {author} {\bibfnamefont
  {L.}~\bibnamefont {Forr{\'o}}}, \bibinfo {author} {\bibfnamefont
  {F.}~\bibnamefont {Scheffold}},\ and\ \bibinfo {author} {\bibfnamefont
  {S.}~\bibnamefont {Jeney}},\ }\href@noop {} {\bibfield  {journal} {\bibinfo
  {journal} {Physical Review E}\ }\textbf {\bibinfo {volume} {90}},\ \bibinfo
  {pages} {060301} (\bibinfo {year} {2014})}\BibitemShut {NoStop}%
\bibitem [{\citenamefont {Jannasch}\ \emph {et~al.}(2011)\citenamefont
  {Jannasch}, \citenamefont {Mahamdeh},\ and\ \citenamefont
  {Sch{\"a}ffer}}]{JannaschMahamdehSchaffer2011}%
  \BibitemOpen
  \bibfield  {author} {\bibinfo {author} {\bibfnamefont {A.}~\bibnamefont
  {Jannasch}}, \bibinfo {author} {\bibfnamefont {M.}~\bibnamefont {Mahamdeh}},\
  and\ \bibinfo {author} {\bibfnamefont {E.}~\bibnamefont {Sch{\"a}ffer}},\
  }\href@noop {} {\bibfield  {journal} {\bibinfo  {journal} {Physical Review
  Letters}\ }\textbf {\bibinfo {volume} {107}},\ \bibinfo {pages} {228301}
  (\bibinfo {year} {2011})}\BibitemShut {NoStop}%
\bibitem [{\citenamefont {Weitz}\ \emph {et~al.}(1989)\citenamefont {Weitz},
  \citenamefont {Pine}, \citenamefont {Pusey},\ and\ \citenamefont
  {Tough}}]{WeitzPinePuseyTough1989}%
  \BibitemOpen
  \bibfield  {author} {\bibinfo {author} {\bibfnamefont {D.~A.}\ \bibnamefont
  {Weitz}}, \bibinfo {author} {\bibfnamefont {D.~J.}\ \bibnamefont {Pine}},
  \bibinfo {author} {\bibfnamefont {P.~N.}\ \bibnamefont {Pusey}},\ and\
  \bibinfo {author} {\bibfnamefont {R.}~\bibnamefont {Tough}},\ }\href@noop {}
  {\bibfield  {journal} {\bibinfo  {journal} {Physical Review Letters}\
  }\textbf {\bibinfo {volume} {63}},\ \bibinfo {pages} {1747} (\bibinfo {year}
  {1989})}\BibitemShut {NoStop}%
\bibitem [{\citenamefont {Segre}\ and\ \citenamefont
  {Pusey}(1996)}]{SegrePusey1996}%
  \BibitemOpen
  \bibfield  {author} {\bibinfo {author} {\bibfnamefont {P.}~\bibnamefont
  {Segre}}\ and\ \bibinfo {author} {\bibfnamefont {P.}~\bibnamefont {Pusey}},\
  }\href@noop {} {\bibfield  {journal} {\bibinfo  {journal} {Physical Review
  Letters}\ }\textbf {\bibinfo {volume} {77}},\ \bibinfo {pages} {771}
  (\bibinfo {year} {1996})}\BibitemShut {NoStop}%
\bibitem [{\citenamefont {Sperl}(2005)}]{Sperl2005}%
  \BibitemOpen
  \bibfield  {author} {\bibinfo {author} {\bibfnamefont {M.}~\bibnamefont
  {Sperl}},\ }\href@noop {} {\bibfield  {journal} {\bibinfo  {journal}
  {Physical Review E}\ }\textbf {\bibinfo {volume} {71}},\ \bibinfo {pages}
  {060401} (\bibinfo {year} {2005})}\BibitemShut {NoStop}%
\bibitem [{\citenamefont {Safdari}\ \emph {et~al.}(2017)\citenamefont
  {Safdari}, \citenamefont {Cherstvy}, \citenamefont {Chechkin}, \citenamefont
  {Bodrova},\ and\ \citenamefont
  {Metzler}}]{SafdariCherstvyChechkinBodrovaMetzler2017}%
  \BibitemOpen
  \bibfield  {author} {\bibinfo {author} {\bibfnamefont {H.}~\bibnamefont
  {Safdari}}, \bibinfo {author} {\bibfnamefont {A.~G.}\ \bibnamefont
  {Cherstvy}}, \bibinfo {author} {\bibfnamefont {A.~V.}\ \bibnamefont
  {Chechkin}}, \bibinfo {author} {\bibfnamefont {A.}~\bibnamefont {Bodrova}},\
  and\ \bibinfo {author} {\bibfnamefont {R.}~\bibnamefont {Metzler}},\
  }\href@noop {} {\bibfield  {journal} {\bibinfo  {journal} {Physical Review
  E}\ }\textbf {\bibinfo {volume} {95}},\ \bibinfo {pages} {012120} (\bibinfo
  {year} {2017})}\BibitemShut {NoStop}%
\bibitem [{\citenamefont {Khan}\ and\ \citenamefont
  {Mason}(2014)}]{KhanMason2014}%
  \BibitemOpen
  \bibfield  {author} {\bibinfo {author} {\bibfnamefont {M.}~\bibnamefont
  {Khan}}\ and\ \bibinfo {author} {\bibfnamefont {T.~G.}\ \bibnamefont
  {Mason}},\ }\href@noop {} {\bibfield  {journal} {\bibinfo  {journal}
  {Physical Review E}\ }\textbf {\bibinfo {volume} {89}},\ \bibinfo {pages}
  {042309} (\bibinfo {year} {2014})}\BibitemShut {NoStop}%
\bibitem [{\citenamefont {Ghosh}\ and\ \citenamefont
  {Krishnamurthy}(2018)}]{GhoshKrishnamurthy2018}%
  \BibitemOpen
  \bibfield  {author} {\bibinfo {author} {\bibfnamefont {K.}~\bibnamefont
  {Ghosh}}\ and\ \bibinfo {author} {\bibfnamefont {C.~V.}\ \bibnamefont
  {Krishnamurthy}},\ }\href@noop {} {\bibfield  {journal} {\bibinfo  {journal}
  {Physical Review E}\ }\textbf {\bibinfo {volume} {98}},\ \bibinfo {pages}
  {052115} (\bibinfo {year} {2018})}\BibitemShut {NoStop}%
\bibitem [{\citenamefont {Kenkre}\ \emph {et~al.}(1981)\citenamefont {Kenkre},
  \citenamefont {K{\"u}hne},\ and\ \citenamefont
  {Reineker}}]{KenkreKuhneReineker1981}%
  \BibitemOpen
  \bibfield  {author} {\bibinfo {author} {\bibfnamefont {V.}~\bibnamefont
  {Kenkre}}, \bibinfo {author} {\bibfnamefont {R.}~\bibnamefont {K{\"u}hne}},\
  and\ \bibinfo {author} {\bibfnamefont {P.}~\bibnamefont {Reineker}},\
  }\href@noop {} {\bibfield  {journal} {\bibinfo  {journal} {Zeitschrift
  f{\"u}r Physik B Condensed Matter}\ }\textbf {\bibinfo {volume} {41}},\
  \bibinfo {pages} {177} (\bibinfo {year} {1981})}\BibitemShut {NoStop}%
\bibitem [{\citenamefont {Bian}\ \emph {et~al.}(2016)\citenamefont {Bian},
  \citenamefont {Kim},\ and\ \citenamefont
  {Karniadakis}}]{BianKimKarniadakis2016}%
  \BibitemOpen
  \bibfield  {author} {\bibinfo {author} {\bibfnamefont {X.}~\bibnamefont
  {Bian}}, \bibinfo {author} {\bibfnamefont {C.}~\bibnamefont {Kim}},\ and\
  \bibinfo {author} {\bibfnamefont {G.~E.}\ \bibnamefont {Karniadakis}},\
  }\href@noop {} {\bibfield  {journal} {\bibinfo  {journal} {Soft Matter}\
  }\textbf {\bibinfo {volume} {12}},\ \bibinfo {pages} {6331} (\bibinfo {year}
  {2016})}\BibitemShut {NoStop}%
\bibitem [{\citenamefont {Giesekus}(1995)}]{Giesekus1995}%
  \BibitemOpen
  \bibfield  {author} {\bibinfo {author} {\bibfnamefont {H.}~\bibnamefont
  {Giesekus}},\ }\href@noop {} {\bibfield  {journal} {\bibinfo  {journal}
  {Rheologica Acta}\ }\textbf {\bibinfo {volume} {34}},\ \bibinfo {pages} {2}
  (\bibinfo {year} {1995})}\BibitemShut {NoStop}%
\bibitem [{\citenamefont {Makris}\ and\ \citenamefont
  {Kampas}(2009)}]{MakrisKampas2009}%
  \BibitemOpen
  \bibfield  {author} {\bibinfo {author} {\bibfnamefont {N.}~\bibnamefont
  {Makris}}\ and\ \bibinfo {author} {\bibfnamefont {G.}~\bibnamefont
  {Kampas}},\ }\href@noop {} {\bibfield  {journal} {\bibinfo  {journal}
  {Rheologica Acta}\ }\textbf {\bibinfo {volume} {48}},\ \bibinfo {pages} {815}
  (\bibinfo {year} {2009})}\BibitemShut {NoStop}%
\bibitem [{\citenamefont {Makris}\ and\ \citenamefont
  {Efthymiou}(2020)}]{MakrisEfthymiou2020}%
  \BibitemOpen
  \bibfield  {author} {\bibinfo {author} {\bibfnamefont {N.}~\bibnamefont
  {Makris}}\ and\ \bibinfo {author} {\bibfnamefont {E.}~\bibnamefont
  {Efthymiou}},\ }\href@noop {} {\bibfield  {journal} {\bibinfo  {journal}
  {Rheologica Acta}\ }\textbf {\bibinfo {volume} {59}},\ \bibinfo {pages} {849}
  (\bibinfo {year} {2020})}\BibitemShut {NoStop}%
\bibitem [{\citenamefont {Clough}\ and\ \citenamefont
  {Penzien}(1970)}]{CloughPenzien1970}%
  \BibitemOpen
  \bibfield  {author} {\bibinfo {author} {\bibfnamefont {R.~W.}\ \bibnamefont
  {Clough}}\ and\ \bibinfo {author} {\bibfnamefont {J.}~\bibnamefont
  {Penzien}},\ }\href@noop {} {\emph {\bibinfo {title} {Dynamics of
  structures}}}\ (\bibinfo  {publisher} {McGraw-Hill},\ \bibinfo {address} {New
  York, NY},\ \bibinfo {year} {1970})\BibitemShut {NoStop}%
\bibitem [{\citenamefont {Harris}\ and\ \citenamefont
  {Crede}(1976)}]{HarrisCrede1976}%
  \BibitemOpen
  \bibfield  {author} {\bibinfo {author} {\bibfnamefont {C.~M.}\ \bibnamefont
  {Harris}}\ and\ \bibinfo {author} {\bibfnamefont {C.~E.}\ \bibnamefont
  {Crede}},\ }\href@noop {} {\emph {\bibinfo {title} {Shock and vibration
  handbook}}},\ \bibinfo {edition} {2nd}\ ed.\ (\bibinfo  {publisher}
  {McGraw-Hill},\ \bibinfo {address} {New York, NY},\ \bibinfo {year}
  {1976})\BibitemShut {NoStop}%
\bibitem [{\citenamefont {Oppenheim}\ and\ \citenamefont
  {Schafer}(1975)}]{OppenheimSchafer1975}%
  \BibitemOpen
  \bibfield  {author} {\bibinfo {author} {\bibfnamefont {A.~V.}\ \bibnamefont
  {Oppenheim}}\ and\ \bibinfo {author} {\bibfnamefont {R.~W.}\ \bibnamefont
  {Schafer}},\ }\href@noop {} {\emph {\bibinfo {title} {Digital signal
  processing}}}\ (\bibinfo  {publisher} {Prentice-Hall, Inc.},\ \bibinfo
  {address} {Englewood Cliffs, NJ},\ \bibinfo {year} {1975})\BibitemShut
  {NoStop}%
\bibitem [{\citenamefont {Reid}(1983)}]{Reid1983}%
  \BibitemOpen
  \bibfield  {author} {\bibinfo {author} {\bibfnamefont {G.~J.}\ \bibnamefont
  {Reid}},\ }\href@noop {} {\bibfield  {journal} {\bibinfo  {journal}
  {McGraw-Hill Series in Electrical Engineering}\ } (\bibinfo {year}
  {1983})}\BibitemShut {NoStop}%
\bibitem [{\citenamefont {Borovi{\v{c}}ka}\ and\ \citenamefont
  {Hansen}(2016)}]{BorovivckaHansen2016}%
  \BibitemOpen
  \bibfield  {author} {\bibinfo {author} {\bibfnamefont {J.}~\bibnamefont
  {Borovi{\v{c}}ka}}\ and\ \bibinfo {author} {\bibfnamefont {L.~P.}\
  \bibnamefont {Hansen}},\ }in\ \href@noop {} {\emph {\bibinfo {booktitle}
  {Handbook of Macroeconomics}}},\ Vol.~\bibinfo {volume} {2}\ (\bibinfo
  {publisher} {Elsevier},\ \bibinfo {year} {2016})\ pp.\ \bibinfo {pages}
  {1641--1696}\BibitemShut {NoStop}%
\bibitem [{\citenamefont {Gurtin}(1982)}]{Gurtin1982}%
  \BibitemOpen
  \bibfield  {author} {\bibinfo {author} {\bibfnamefont {M.~E.}\ \bibnamefont
  {Gurtin}},\ }\href@noop {} {\emph {\bibinfo {title} {An introduction to
  continuum mechanics}}}\ (\bibinfo  {publisher} {Academic press},\ \bibinfo
  {year} {1982})\BibitemShut {NoStop}%
\bibitem [{\citenamefont {Bird}\ \emph {et~al.}(1987)\citenamefont {Bird},
  \citenamefont {Armstrong},\ and\ \citenamefont
  {Hassager}}]{BirdArmstrongHassager1987}%
  \BibitemOpen
  \bibfield  {author} {\bibinfo {author} {\bibfnamefont {R.~B.}\ \bibnamefont
  {Bird}}, \bibinfo {author} {\bibfnamefont {R.~C.}\ \bibnamefont
  {Armstrong}},\ and\ \bibinfo {author} {\bibfnamefont {O.}~\bibnamefont
  {Hassager}},\ }\href@noop {} {\emph {\bibinfo {title} {Dynamics of polymeric
  liquids. Vol. 1: Fluid mechanics}}},\ \bibinfo {edition} {2nd}\ ed.\
  (\bibinfo  {publisher} {Wiley},\ \bibinfo {address} {New York, NY},\ \bibinfo
  {year} {1987})\BibitemShut {NoStop}%
\bibitem [{\citenamefont {Liverpool}\ and\ \citenamefont
  {MacKintosh}(2005)}]{LiverpoolMacKintosh2005}%
  \BibitemOpen
  \bibfield  {author} {\bibinfo {author} {\bibfnamefont {T.}~\bibnamefont
  {Liverpool}}\ and\ \bibinfo {author} {\bibfnamefont {F.}~\bibnamefont
  {MacKintosh}},\ }\href@noop {} {\bibfield  {journal} {\bibinfo  {journal}
  {Physical review letters}\ }\textbf {\bibinfo {volume} {95}},\ \bibinfo
  {pages} {208303} (\bibinfo {year} {2005})}\BibitemShut {NoStop}%
\bibitem [{\citenamefont {Atakhorrami}\ \emph {et~al.}(2008)\citenamefont
  {Atakhorrami}, \citenamefont {Mizuno}, \citenamefont {Koenderink},
  \citenamefont {Liverpool}, \citenamefont {MacKintosh},\ and\ \citenamefont
  {Schmidt}}]{AtakhorramiMizunoKoenderinkLiverpoolMacKintoshSchmidt2008}%
  \BibitemOpen
  \bibfield  {author} {\bibinfo {author} {\bibfnamefont {M.}~\bibnamefont
  {Atakhorrami}}, \bibinfo {author} {\bibfnamefont {D.}~\bibnamefont {Mizuno}},
  \bibinfo {author} {\bibfnamefont {G.}~\bibnamefont {Koenderink}}, \bibinfo
  {author} {\bibfnamefont {T.}~\bibnamefont {Liverpool}}, \bibinfo {author}
  {\bibfnamefont {F.}~\bibnamefont {MacKintosh}},\ and\ \bibinfo {author}
  {\bibfnamefont {C.}~\bibnamefont {Schmidt}},\ }\href@noop {} {\bibfield
  {journal} {\bibinfo  {journal} {Physical Review E}\ }\textbf {\bibinfo
  {volume} {77}},\ \bibinfo {pages} {061508} (\bibinfo {year}
  {2008})}\BibitemShut {NoStop}%
\end{thebibliography}%

\end{document}